\documentclass[apj]{emulateapj}
\usepackage{graphicx}
\usepackage{epstopdf}
\usepackage{amsmath}
\usepackage{amssymb}
\shorttitle{Mass accretion rate}
\shortauthors{Bu \& Yang}
\begin{document}
\title{Quenching black hole accretion by active galactic nuclei feedback}
\author{De-Fu Bu\altaffilmark{1}, Xiao-Hong Yang\altaffilmark{2}}

\altaffiltext{1}{Key Laboratory for Research in Galaxies and Cosmology, Shanghai Astronomical Observatory, Chinese Academy of Sciences, 80 Nandan Road, Shanghai 200030, China; dfbu@shao.ac.cn }
\altaffiltext{2}{Department of physics, Chongqing University, Chongqing, 400044; yangxh@cqu.edu.cn}

%\date{Accepted 1988 December 15. Received 1988 December 14; in original form 1988 October 11}

%\pagerange{\pageref{firstpage}--\pageref{lastpage}} \pubyear{2002}

%\maketitle

%\label{firstpage}
%
\begin{abstract}
Observations of many dim galactic nuclei in local universe give good estimation of gas density and temperature at Bondi radius. If we assume the black hole accretes at Bondi accretion rate and radiates at efficiency of low-luminosity hot accretion flow, the predicted nuclei luminosity can be significantly higher than observations. Therefore, the real black hole mass accretion rate in these sources may be significantly smaller than the Bondi value. Active galactic nuclei (AGN) feedback may be responsible for decreasing black hole accretion rate to values much smaller than Bondi rate. We perform two-dimensional simulations of low angular momentum accretion flow at parsec and sub-parsec scales around low-luminosity active galactic nuclei (LLAGNs). We take into account the radiation and wind feedbacks of the LLAGN. The cross section of particle-particle interaction can be several orders of magnitude larger than that of photon-particle interaction. Therefore, we find that for the LLAGNs, effects of radiation feedback in decreasing black hole accretion rate are small. However, wind feedback can effectively decrease the black hole mass accretion rate. Due to the decrease of accretion rate, the black hole luminosity can be decreased by a factor of $\sim 33-400$. The results may be useful to explain why many galactic nuclei in the local universe are so dim.
\end{abstract}


\section{Introduction}
Observations show that most nearby galactic nuclei are very dim. There are some examples. The luminosity of the back hole accretion flow in our Galactic center (Sgr A*) is $L \sim 10^{-9} L_{\rm Edd}$ ($L_{\rm Edd}$ is Eddington luminosity, Yuan et al. 2003). In some nearby elliptical galaxies including NGC 1399, 4636 and 4472, the black hole accretion flow luminosity is $L < 10^{-8} L_{\rm Edd}$ (Loewenstein et al. 2001).

Recently, Pellegrini (2005) collects a sample of nearby galaxies of \emph{Chandra} observations. For these galaxies, the black hole mass ($M_{\rm BH}$), the X-ray luminosity of the nuclei, and the gas density and temperature at the Bondi radius are accurately estimated. The black holes in these galaxies are very dim with $L<10^{-4} L_{\rm Edd}$.
The Bondi accretion rate ($\dot M_{\rm B}$) of these sources can be very easily obtained based on the gas density and temperature at Bondi radius. If a standard thin disk (Shakura \& Sunyaev 1973) operates in these sources, one would expect the luminosity of the black hole should be $L=\eta \dot M_{\rm B} c^2$. $\eta$ is the radiative efficiency which depends on the black hole spin. For a moderately spinning black hole, $\eta=0.1$ (Wu, Lu, Zhang et al. 2013). Pellegrini (2005) finds that the luminosity of these nuclei is significantly smaller than $0.1 \dot M_{\rm B} c^2$. Therefore, the standard thin disk model can not be applied to these sources.

Observations of black hole X-ray binaries (BHBs) find that the black hole always transients from soft to hard states at $2\%L_{\rm Edd}$ (McClintock \& Remillard 2006). The black hole accretion physics does not depend on black hole mass. Luminous AGNs correspond to soft state of BHBs. The accretion disk model of Luminous AGNs is standard thin disk. The LLAGNs correspond to the hard state of BHBs (Ho 2008). The accretion disk model of LLAGNs is hot accretion flow (see Yuan \& Narayan 2014 for reviews). Yuan \& Li (2011) found that the boundary between LLAGNs and luminous AGNs is $2\%L_{\rm Edd}$. The nuclei luminosity of galaxies collected by Pellegrini (2005) $L<10^{-4} L_{\rm Edd}$. Therefore, the accretion flows in these sources should be radiatively inefficient hot accretion flow.

The earliest version of hot accretion flow is advection dominated accretion flow (ADAF; Narayan \& Yi 1994; 1995). In ADAF solution, outflow is absent, the accretion rate is a constant with radius. According to the ADAF model, the mass accretion rate at the black hole horizon equals to $\dot M_{\rm B}$. The radiative efficiency of ADAF is much smaller than 0.1. Pellegrini (2005) finds that the predicted luminosity by the ADAF model can be consistent with that of observations of a few nearby galactic nuclei. However, for many other galactic nuclei, the ADAF predicted luminosity is still too higher.

There are also two other types of radiatively inefficient hot accretion flow. The first one is the adiabatic inflow-outflow solution (ADIOS; Blandford \& Begelman 1999, 2004; Begelman 2012). The ADIOS model assumes that outflow can be launched at any radii. Due to the presence of outflow, the mass accretion rate is not a constant with radius. Mass accretion rate decreases towards the black hole. According to the ADIOS model, the mass accretion rate at black hole horizon can be significantly smaller than $\dot M_{\rm B}$. The second one is the convection-dominated accretion flow (CDAF; Narayan et al. 2000; Quataert \& Gruzinov 2000; Inayoshi et al. 2018). The CDAF model assumes that convective motions dominate the dynamics of accretion flow. With accretion, more and more gas will be locked in convective eddies. The mass accretion rate decreases towards the black hole. CDAF model also predicts that the mass accretion rate at black hole horizon can be significantly smaller than $\dot M_{\rm B}$. Therefore, both the ADIOS and CDAF models may be able to explain the low emission level of the observed nearby galactic nuclei.

We note that both the ADIOS and CDAF models can only be applied to high angular momentum accretion gas. In ADIOS and CDAF models, at any radius, the gas angular momentum is comparable to the local Keplerian angular momentum. We define ``circularization radius" ($r_c$) as follows. The specific angular momentum of gas at Bondi radius is equal to the Keplerican angular momentum at $r_c$. If $r_c$ is significantly smaller than Bondi radius, ADIOS and CDAF models can not be applicable. Then a question is that if ADIOS and CDAF models are not applicable, how to explain the low emission level of the observed nearby galactic nuclei?

AGN feedback (Fabian 2012) is a possible mechanism to reduce the black hole accretion rate. The outputs of an AGN include radiation, wind and jet. The AGN emitted photons can heat or cool the gas at large radii by Compton heating/cooling (Ciotti \& Ostriker 1997, 2001, 2007; Ciotti et al. 2009). The heating/cooling rate depends on nuclei luminosity, Compton temperature of AGN emitted photons and accretion gas temperature at large radii. Recent numerical simulations of LLAGN with radiative feedback find that the AGN emitted photons can compton heat the gas around the Bondi radius. The gas temperature around Bondi radius can be much higher than the local virial temperature. Thus, outflows can be launched around Bondi radius (Yang \& Bu 2018; Bu \& Yang 2018). Radiation pressure can also directly push gas away. For example, for a quasar, the not fully ionized gas can absorb the UV photons. Consequently, line force will be exerted on the accreting gas. Line force can exceed black hole gravity significantly. Strong outflow can be launched by line force (e.g., Proga et al. 2000; Murray et al. 1995; Murray \& Chiang 1997; Liu et al. 2013; Nomura et al. 2016; Nomura \& Ohsuga 2017). Recently, we find that if a hot corona exists above a standard thin disk, the radiation force due to Thomson scattering can also drive wind from the hot corona (Yang et al. 2018).

AGN wind can also effectively interact with gas surrounding the AGN (e.g., Ostriker et al. 2010; Weinberger et al. 2017, 2018; Gan et al. 2014; Yuan et al. 2018). Wind can directly blow the gas surrounding AGN away, which will result in the decrease of black hole accretion rate.

In this paper, we study how the slowly rotating gas falls from the Bondi radius to the black hole. Our computational domain covers a region from 500 $r_s$ ($r_s$ is Schwarzschild radius) to the region beyond the Bondi radius. We focus on LLAGNs whose luminosity is smaller than $2\%L_{\rm Edd}$. We define LLAGN to be the accretion flow inside the inner boundary of the simulation domain. In this sense, the LLAGN is not resolved. We take into account the radiation and wind feedbacks from the LLAGN. For the radiation feedback, we consider the Compton heating/cooling effects. Also, the radiation pressure due to the Thomson scattering is taken into account. The feedback by jet is not taken into account. The reason is as follows. Jet is well collimated with very small opening angle. In this case, jet may just pierce through the galaxy and have negligible effects on the gas close to the galactic center. We note that jet may be important at galaxy cluster scales (Guo 2016; Guo et al. 2018). Our purpose is to study the effects of LLAGN feedback on decreasing black hole accretion rate. This may be useful to explain the low emission level of the observed nearby galactic nuclei.

As mentioned above, recently, there are works studying low-angular momentum accretion flow around LLAGN (Bu \& Yang 2018; Yang \& Bu 2018). In these two works, only the radiation feedback of the LLAGN is taken into account. In the present paper, in addition to radiation feedback, we also have wind feedback. We find that wind feedback is more effective in decreasing the black hole accretion rate.

We organize our paper as follows. In section 2, we introduce the numerical method and physical assumptions. In section 3, we present our results; Section 4 is devoted to summary and discussion.

\section{Numerical method }
In this paper, we set the black hole mass $M_{\rm BH}=10^8M_{\odot}$, $M_{\odot}$ is solar mass.
We perform two-dimensional numerical simulations using the ZEUS-MP code (Hayes et al. 2006). In spherical coordinates ($r,\theta,\phi$), we solve the equations below:

\begin{equation}
 \frac{d\rho}{dt} + \rho \nabla \cdot {\bf v} = 0,
\end{equation}
\begin{equation}
 \rho \frac{d{\bf v}}{dt} = -\nabla p - \rho \nabla \Phi + \rho \bf{F_{rad}}
\end{equation}
\begin{equation}
 \rho \frac{d(e/\rho)}{dt} = -p\nabla \cdot {\bf v} + Sc - Br
\end{equation}
$\rho$, $\bf v$ and $e$ are density, velocity and internal energy, respectively. We adopt ideal gas equation $p=(\gamma-1)e$ and set $\gamma=5/3$. The black hole potential is $\Phi_{\rm BH}=-GM_{\rm BH}/(r-r_s)$, with $G$ being gravitational constant. $Br=3.8\times 10^{-27}n^2\sqrt{T}$ is the bremsstrahlung cooling. $n=\rho/(\mu m_p)$ is number density of gas, with $\mu$ and $m_p$ being mean molecular weight and proton mass, respectively. We set $\mu=0.5$. $T$ is the temperature of the accreting gas.

We summarize the implementations of physics in the simulations in Table 1. Below, we introduce in details these information.

\subsection{Radiation feedback}
The radiation of the central LLAGN can have a radiation pressure on the accretion gas. In Equation (2), ${\rho F_{rad}}=\frac{\chi}{c}\frac{L}{4\pi r^2}$ is the radiation pressure due to Compton scattering. $L$ is the luminosity of LLAGN. The calculation of luminosity of the central LLAGN will be introduced below. $\chi$ is the Compton scattering opacity. In addition to radiation pressure, the photons emitted by the LLAGN can also Compton heating/cooling the accretion gas. The Compton heating/cooling rate in Equation (3) is:
\begin{equation}
Sc=4.1\times10^{-35}n^2(T_X-T)\xi
\end{equation}
$T_X$ is the Compton temperature of the photons emitted by the central LLAGN. Xie et al. (2017) find that for a LLAGN, the Compton temperature $T_X$ of the photons is $\sim 10^8$K. In this paper, we set $T_X=10^8$K. $\xi=L e^{-\tau}/nr^2$ is the ionization parameter, where $\tau=\int_0^r \rho \kappa_X dr$ is the X-ray scattering optical depth. We set $\kappa_X=0.4 {\rm cm^2 g^{-1}}$.

\subsection{Wind feedback}
Winds are frequently observed through blue shifted absorption lines in luminous AGNs (e.g., Crenshaw et al. 2003; Tombesi et al. 2010, 2014; King \& Pounds 2015; Liu et al. 2015; Gofford et al. 2015) and soft state of BHBs (e.g., Neilsen \& Homan 2012; Homan et al. 2016; D\'{i}az Trigo \& Boirin 2016).

The observations of wind from LLAGNs and hard state of BHBs are very few. The reason may be that hot accretion flow is fully ionized. Therefore, it is hard to detect absorption lines. In recent years, we have some indirect evidences showing that wind can also be generated in hot accretion flow (e.g., Crenshaw \& Kramemer 2012; Wang et al. 2013; Cheung et al. 2016; Homan et al. 2016). Recently, the detailed properties of wind from hot accretion flow are studied by numerical simulations (e.g., Tchekhovskoy et al. 2011; Yuan et al. 2012, 2015; Narayan et al. 2012; Li et al. 2013; see also Moller \& Sadowski 2015) and analytical works (e.g., Cao 2011; Wu, Cao, Ho et al. 2013; Gu 2015).

Simulations of hot accretion flow with large angular momentum (e.g., Yuan et al. 2015) find that wind can be generated outside $10r_s$. Large angular momentum means that the gas angular momentum is comparable to the local Keplerian angular momentum. The absence of wind inside $10r_s$ is due to the very strong gravity very close to the black hole. The wind is generated by the combination of gas pressure gradient, magnetic pressure gradient and centrifugal forces. Due to the presence of wind, the mass accretion rate decreases inwards. The black hole mass accretion rate can be described as follows (e.g., Yuan et al. 2012, 2015):
\begin{equation}
\dot M_{\rm BH}=\dot M_{\rm {R_H}}\left(\frac{10r_s}{R_{\rm H}}\right)^{0.5}
\end{equation}
$R_{\rm H}$ is the outer boundary of the hot accretion flow; $\dot M_{\rm R_H}$ is the mass accretion rate at $R_{\rm H}$. As mentioned in Section 1, in this paper, we study slowly rotating accretion flow. We set the circularization radius $r_c=350r_s$, which is much smaller than the inner radial boundary of the computational domain. When gas flows through the inner boundary, we assume that gas will freely fall to $r_c$. When gas arrives at $r_c$, a viscous hot accretion flow will form. Wind will be generated from the viscous hot accretion flow. Therefore, the outer boundary of the hot accretion flow in Equation (5) is $R_{\rm H}=r_c$. The mass accretion rate at $R_{\rm H}$ is equal to the mass accretion rate calculated at the inner radial boundary of our simulation domain ($\dot M_{\rm in}$). The mass flux of wind generated in the viscous hot accretion flow is
\begin{equation}
\dot M_{\rm wind}=\dot M_{\rm in}-\dot M_{\rm BH}
\end{equation}

Yuan et al. (2015) have shown that the radial velocity of wind is only a function of wind launching radius:
\begin{equation}
 v_{\rm wind,r}=0.21v_k(R_{\rm wind})
\end{equation}
$R_{\rm wind}$ is wind launching radius, $v_k(R_{\rm wind})$ is the Keplerian velocity at wind launching radius. Once wind is launched, its radial velocity will not change with the outward propagation of wind. This is because that acceleration forces are always exerted on wind with the outwards moving of wind. Works done by these forces can compensate the increase of gravitational energy of wind (Yuan et al. 2015). As mentioned above, wind can be generated by hot accretion flow in the region $10r_s<r<R_{\rm H}$. We can calculate the mass flux weighted radial velocity of wind as follows:
%\begin{equation}
%\overline{Vr}_{\rm wind}=\frac{\int_{10r_s}^{\rm R_H} \frac {d \left(\dot M_{\rm {R_H}}\left(\frac{r}{R_{\rm H}}\right)^{0.5} \right)}{dr} Vr_{\rm wind} dr}{\int_{10r_s}^{\rm R_H} \frac {d \left(\dot M_{\rm {R_H}}\left(\frac{r}{R_{\rm H}}\right)^{0.5} \right)}{dr}dr}
%\end{equation}
\begin{equation}
\overline{v}_{\rm wind,r}=\frac{\int_{10r_s}^{\rm R_H} \frac {d \left(\dot M_{\rm {wind}} (r) \right)}{dr} v_{\rm wind,r} dr}{\int_{10r_s}^{\rm R_H} \frac {d \left(\dot M_{\rm {wind}}(r) \right)}{dr}dr}
\end{equation}
According to Equation (6), $d \dot M_{\rm wind}/dr= d \dot M_{\rm in}/dr$, with $\dot M_{\rm in}(r)= \dot M_{\rm R_H}\left(\frac{r}{R_{\rm H}}\right)^{0.5}$ (Yuan et al. 2015). We find that with $R_H=r_c=350r_s$, the mass flux weighted wind radial velocity is $0.42v_k(350r_s)$. $v_k(350r_s)$ is the Keplerian angular velocity at $350r_s$.
Yuan et al. (2015) find that $v_\theta$ of wind is negligibly small than the radial velocity. Thus, we set $v_\theta$ of wind to be zero. The rotational velocity of wind is found to be just slightly smaller than Keplerian velocity (Yuan et al. 2012, 2015). In this paper, we set the rotational velocity of wind to be:
\begin{equation}
V_{\rm \phi, wind}=0.9v_k(R_{\rm H})
\end{equation}

Yuan et al. (2015) find that the internal energy of wind is about 0.6 times the gravitational energy. In this paper, we set wind internal energy:
\begin{equation}
e_{\rm wind}=\frac{0.6GM_{\rm BH}}{R_{\rm H}}
\end{equation}

Detailed analysis of wind in Yuan et al. (2015) show that wind mass flux is distributed within $\theta \sim 30^\circ-70^\circ$ and $\theta \sim 110^\circ-150^\circ$. In this paper, we only simulate the region above the equatorial plane. Therefore, we inject wind at the inner radial boundary in the region $30^\circ <\theta <70^\circ$. The injected mass flux of wind is given by Equation (6). In the region $30^\circ <\theta <70^\circ$, we assume that the wind mass flux is independent of $\theta$.

\subsection{Luminosity of central LLAGN}
The luminosity of the central LLAGN depends on the black hole mass accretion rate, radiative efficiency of hot accretion flow. The black hole accretion rate is calculated by Equation (5). Xie \& Yuan (2012) find that the radiative efficiency depends on black hole accretion rate and the parameter $\delta$ which describes the fraction of the direct viscous heating to electrons. The radiative efficiency is as follows (Xie \& Yuan 2012):
\begin{equation}
\epsilon(\dot{M}_{\text{BH}})=\epsilon_0(\frac{100\dot{M}_{\text{BH}}}{\dot{M}_{\text{Edd}}})^a,
\end{equation}
$\dot{M}_{\text{Edd}}=L_{\rm Edd}/0.1c^2$ is the Eddington accretion rate; $\epsilon_0$ and $a$ for the case of $\delta=0.5$ can be described as:
\begin{equation}
(\epsilon_0,a) = \left\{ \begin{array}{ll}
(1.58,0.65) & \textrm{if } \frac{\dot{M}_{\text{BH}}}{\dot{M}_{\text{Edd}}}\lesssim2.9\times10^{-5};\\
(0.055,0.076) & \textrm{if } 2.9\times10^{-5}<\frac{\dot{M}_{\text{BH}}}{\dot{M}_{\text{Edd}}}\lesssim3.3\times10^{-3};\\
(0.17,1.12) & \textrm{if } 3.3\times10^{-3}<
\frac{\dot{M}_{\text{BH}}}{\dot{M}_{\text{Edd}}}\lesssim5.3\times10^{-3}.
\end{array} \right.
\end{equation}
When $\frac{\dot{M}_{\text{BH}}}{\dot{M}_{\text{Edd}}}>5.3\times10^{-3}$,
the radiative efficiency $\epsilon(\dot{M}_{\text{BH}})$ is simply
set to be 0.1.

\subsection{Initial and boundary conditions}

In radial direction, our computational domain is $500r_s\leq r \leq 10^6r_s$. In $\theta$ direction, we have $0 \leq \theta \leq \pi/2$. Totally, we have $140\times 80$ grids. In $r$ direction, the grids are logarithmically spaced. In $\theta$ direction, grids are uniformly spaced.
Initially, in the whole computational domain, gas has uniform density ($\rho_0$) and temperature ($T_0$). The specific angular momentum of gas is equal to the Keplerian angular momentum at $r_c=350r_{\text{s}}$.

In the models with wind feedback, at the inner radial boundary, in the region $30^\circ<\theta<70^\circ$, wind is injected into the computational domain. The density, velocity and internal energy of injected wind are set according to Equations (6), (8)-(10). In the regions $0^\circ <\theta\leq30^\circ$ and $70^\circ \leq \theta\leq 90^\circ$, we use outflow boundary conditions. For the outflow boundary conditions, gas is not allowed to flow into the computational domain. The physical variables in the ghost zones are set to be equal to those in the fist active zone. In the models without wind feedback, at the inner radial boundary, in the whole $\theta$ angle ($0^\circ \leq \theta \leq 90^\circ$), we use outflow boundary conditions.

We set the outer radial boundary conditions as follows. If the radial velocity at the last active zone at a fixed $\theta$ angle is negative, at this $\theta$ angle, we inject gas into the computational domain. The density, temperature, specific angular momentum of the injected gas are equal to those for the initial condition.  If the radial velocity at the last active zone at a fixed $\theta$ angle is positive, at this $\theta$ angle, we use outflow boundary conditions. Gas is not allowed to flow into the computational domain.

At the rotational axis, axis-of-symmetry boundary conditions are applied. We use reflecting boundary conditions at
the equatorial plane.
\begin{table*} \caption{Implementation of physics and assumptions }
\setlength{\tabcolsep}{4mm}{
\begin{tabular}{ccc}
\hline \hline
  &      {\bf Implementation of wind feedback }             &    \\
\hline
Mass flux of wind for feedback & Equations (5) and (6) & Ref: Yuan et al. (2015) \\
Velocities of wind for feedback & Equations (7), (8), and (9) & Ref: Yuan et al. (2015) \\
Internal energy of wind for feedback & Equation (10) & Ref: Yuan et al. (2015) \\
Injection angle of wind at inner radial boundary & $30^\circ<\theta<70^\circ$ & Ref: Yuan et al. (2015)\\
\hline\hline
&      {\bf Implementation of radiation feedback }             &    \\
\hline
Compton temperature of X-ray photons & $10^8$ K & Ref: Xie et al. (2017) \\
Radiative efficiency & Equation (12) & Ref: Xie \& Yuan (2012) \\
\hline\hline
&      {\bf Initial conditions and other assumptions }             &    \\
\hline
Initial gas temperature & $7\times 10^6$ K &  \\
Initial gas density & $\rho_0=10^{-26}-10^{-22}{\rm g/cm^3}$  & \\
Initial gas angular momentum & $\sqrt{GM_{\rm BH}350r_s}$ & \\

\hline\noalign{\smallskip}
\end{tabular}}
\end{table*}

\begin{table*} \caption{Simulation parameters and results }
%\begin{supertabular}{cccccc}
\setlength{\tabcolsep}{4mm}{
\begin{tabular}{ccccccc}
\hline \hline
 Model & radiation feedback & wind feedback & $\rho_0$  &  $\dot M(r_{\rm in})/\dot M_{\rm B}$ & $L/L_{\rm Edd}$ & $P_{\rm K}(r_{\rm out})$ \\

  &   &  & ($10^{-24}\text{g cm}^{-3}$) &          &  &($L_{\rm Edd}$)   \\
(1) & (2)             & (3)                         &  (4)      &     (5)          &        (6)   & (7)   \\

\hline\noalign{\smallskip}
noFB26   & OFF & OFF & 0.01&  0.1 & $1.2 \times 10^{-10} $ & $1.4 \times 10^{-11}$  \\
radFB26  & ON  & OFF & 0.01&  0.11 & $1.3 \times 10^{-10} $ & $1.5 \times 10^{-11}$  \\
windFB26 & OFF & ON  & 0.01&  $3.9 \times 10^{-2}$ & $3.3\times 10^{-11}$ & $5.4 \times 10^{-10}$ \\
fullFB26 & ON  & ON  & 0.01&  $2.8 \times 10^{-2}$ & $2.1\times 10^{-11}$ & $3.5 \times 10^{-10}$ \\
noFB24   & OFF & OFF & 1   &  0.12 & $3.9 \times 10^{-7} $ & $1.5 \times 10^{-9}$ \\
radFB24  & ON  & OFF & 1   &  0.12 & $3.8 \times 10^{-7} $ & $1.49 \times 10^{-9}$\\
windFB24 & OFF & ON  & 1   &  $3.7 \times 10^{-2}$ & $6.2\times 10^{-8}$ & $4.9 \times 10^{-8}$ \\
fullFB24 & ON  & ON  & 1   &  $1.9 \times 10^{-2}$ & $2.4\times 10^{-8}$ & $2.2 \times 10^{-8}$ \\
noFB22   & OFF & OFF & 100 &  1.05                & $1.5 \times 10^{-2}$ & 0 \\
radFB22  & ON  & OFF & 100 &  0.42                & $5.8 \times 10^{-3}$ & $2.7 \times 10^{-9}$ \\
windFB22 & OFF & ON  & 100 &  $1.5 \times10^{-2}$ & $ 10^{-4}$ & $6.4 \times 10^{-6}$  \\
fullFB22 & ON  & ON  & 100 &  $6.2 \times10^{-2}$ & $4.5 \times 10^{-4}$ & $4.3 \times 10^{-5}$ \\

\hline\noalign{\smallskip}
\end{tabular}}
%\end{supertabular}

Note: Col. 1: model names. Cols 4: the density for initial condition. Col. 5: time-averaged mass accretion rate (in unit of Bondi rate) measured at the inner boundary of the simulation domain. Col. 6: time-averaged luminosity of the black hole. Col. 7: time-averaged mechanical energy flux (in unit of Eddington luminosity) of outflow measured at the outer boundary. For models with $\rho_0=10^{-24}\text{g cm}^{-3}$ and $\rho_0=10^{-26}\text{g cm}^{-3}$, we do the time-average from $t=0$ to $2\times 10^5$ year. For models with $\rho_0=10^{-22}\text{g cm}^{-3}$, we do the time-average from $t=1.2\times 10^5$ to $2.5 \times 10^5$ year.

\end{table*}

\begin{figure*}
\begin{center}
\includegraphics[scale=0.5]{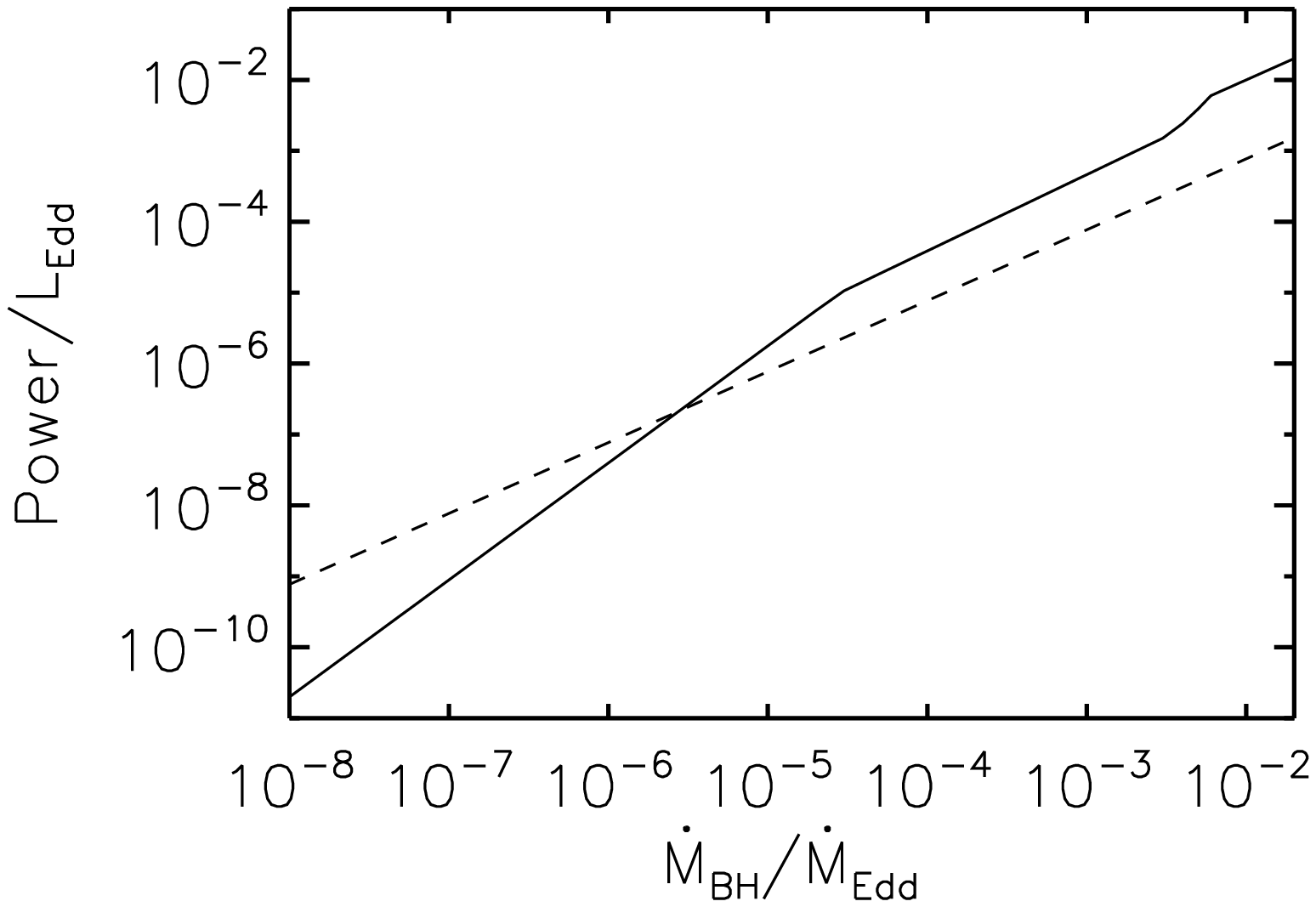}\hspace*{0.7cm}
\includegraphics[scale=0.5]{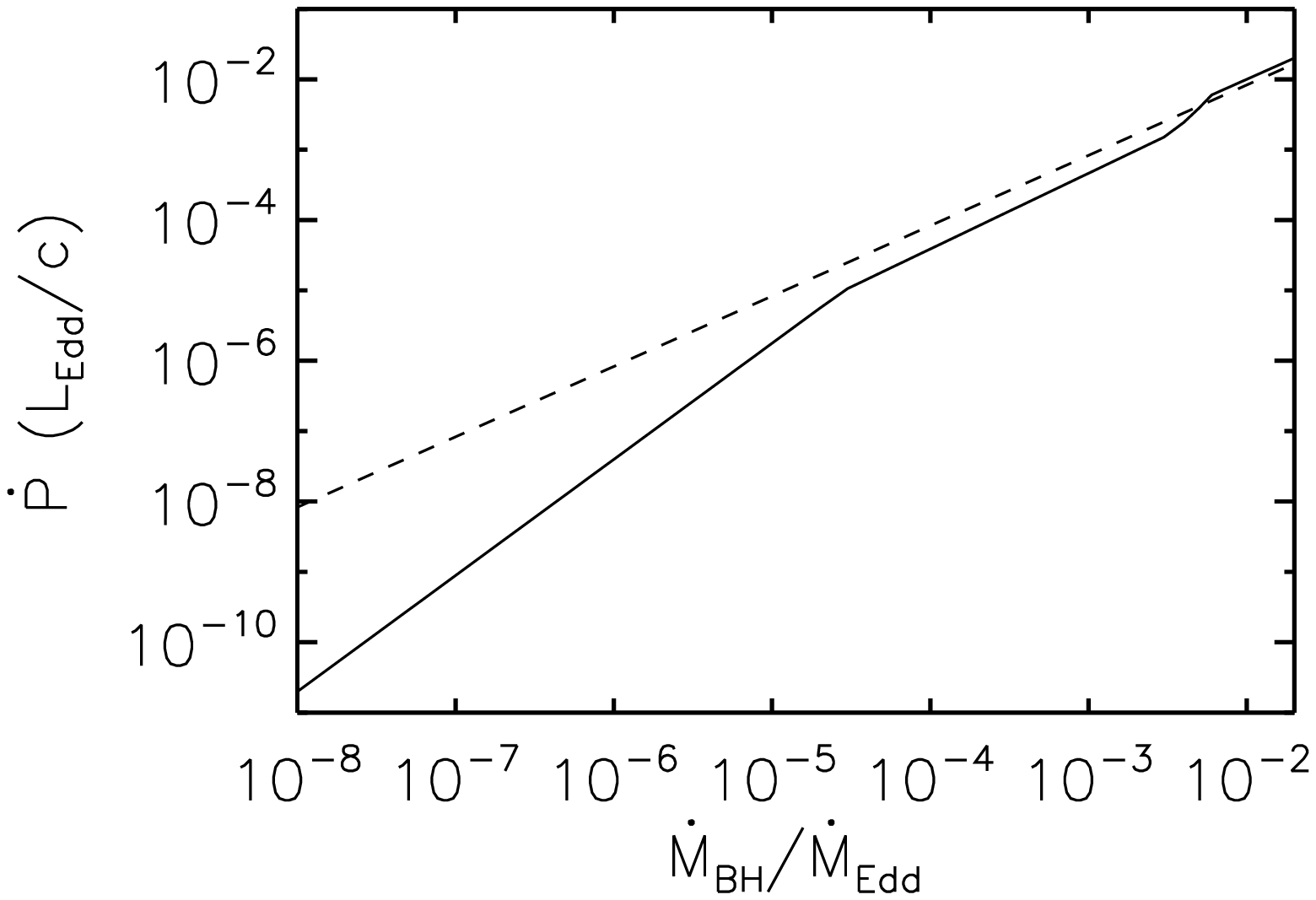}\hspace*{0.7cm}
\hspace*{0.5cm} \caption{Left panel: powers of wind (dashed line) and radiation (solid line) generated from the central LLAGN versus black hole mass accretion rate. Power is in unit of Eddington luminosity ($L_{\rm Edd}$). Black hole mass accretion rate is in unit of Eddington accretion rate ($\dot M_{\rm Edd}=10L_{\rm Edd}/c^2$). Right panel: momentum fluxes of wind (dashed line) and radiation (solid line) generated from the central LLAGN versus black hole mass accretion rate. The momentum flux is in unit of $L_{\rm Edd}/c$. \label{Fig:power}}
\end{center}
\end{figure*}

\section{Results}
In the following of this paper, we use ``wind" to denote the mass output of the central LLAGN, which is injected into the computational domain at the inner radial boundary (see Section 2.2) in models with wind feedback. We use ``outflow" to denote the outward moving gas in the computational domain of our simulations.

Properties of wind and radiation generated from the central LLAGN are introduced in Section 2. We now compare the powers and momentum fluxes carried by wind and radiation.

The power of wind includes kinetic power and thermal power,
\begin{equation}
\dot E_{\rm wind}=\dot M_{\rm wind} (\frac{1}{2}v_{\rm wind}^2+e_{\rm wind}/\rho_{\rm wind})
\end{equation}
For the momentum flux of wind, we only consider the radial momentum flux,
\begin{equation}
\dot P_{\rm wind}=\dot M_{\rm wind} v_{\rm wind,r}
\end{equation}
The momentum flux of radiation from the central LLAGN is
\begin{equation}
\dot P_{\rm rad}=L/c
\end{equation}

The power and momentum flux carried by wind and radiation are shown in Figure \ref{Fig:power}. From this figure, we see that when $\dot M_{\rm BH}/\dot M_{\rm Edd}<3\times 10^{-6}$, wind power dominates radiation power. When $\dot M_{\rm BH}/\dot M_{\rm Edd}>3\times 10^{-6}$, radiation power is much larger than wind power. The interaction efficiency of wind and radiation with the accretion gas at large radii does not only depend on their powers. The reason is that the cross section of proton-particle and particle-particle interactions differs by several orders of magnitude (Yuan et al. 2018). The cross section of photon-particle interaction is the Thompson scattering cross section, $\sigma_T=6.65\times 10^{-25} {\rm cm^2}$. The cross section of Coulomb particle-particle interaction is $\sigma_C\sim\pi e^4/k^2T^2$, where $e$ is electron charge, $k$ is Boltzmann constant. The distances taken for wind and photons to convert their energy and momentum to accretion gas at large radii can be quite different. Following Yuan et al. (2018), we define the ``typical length scale of feedback" for wind ($l_{\rm wind}$) and radiation ($l_{\rm rad}$). They are the distances taken for winds and photons to convert a significant fraction of their energy and momentum to the gas. The length scale of feedback is the mean free path of photons and wind particles. The length scale of feedback is the distance for optical depth $\sim 1$. The typical length scale of radiation feedback is $\sigma_T \rho l_{\rm rad}/m_p=1$. Therefore, we have
\begin{equation}
l_{\rm rad}= m_p/\rho\sigma_T=2.5\times 10^{24} \rho_{-24}^{-1} {\rm cm}  \sim 10^6 \rho_{-24}^{-1} {\rm pc}
\end{equation}
where, $\rho_{-24}=\rho/10^{-24} {\rm g\cdot cm^{-3}} $. For a black hole with $10^8M_\odot$, $l_{\rm rad} \sim 10^{11} \rho_{-24}^{-1} r_s$.
For wind feedback, the typical length scale of feedback is ,
\begin{equation}
l_{\rm wind}= m_p/\rho\sigma_C=1.7\times 10^{19} \rho_{-24}^{-1} T_7^2 {\rm cm} = 5\rho_{-24}^{-1} T_7^2 {\rm pc}
\end{equation}
where $T_7=T/10^7K$.
For $M_{\rm BH}=10^8M_\odot$, we have $l_{\rm wind}=5\times 10^5 \rho_{-24}^{-1} T_7^2 r_s $.

We list all the models in Table 2. The gas temperature at the Bondi radius for the sources collected in Pellegrini (2005) is in the range $2 \times 10^6K \sim 10^7K$. In all the models, we set the initial gas temperature $T_0=7\times 10^6 K$.

%\subsection{Models with $\rho_0=10^{-26}{\rm g/cm^3}$ }

\subsection{Models with $\rho_0=10^{-24}{\rm g/cm^3}$}
\begin{figure}
\begin{center}
\includegraphics[scale=0.5]{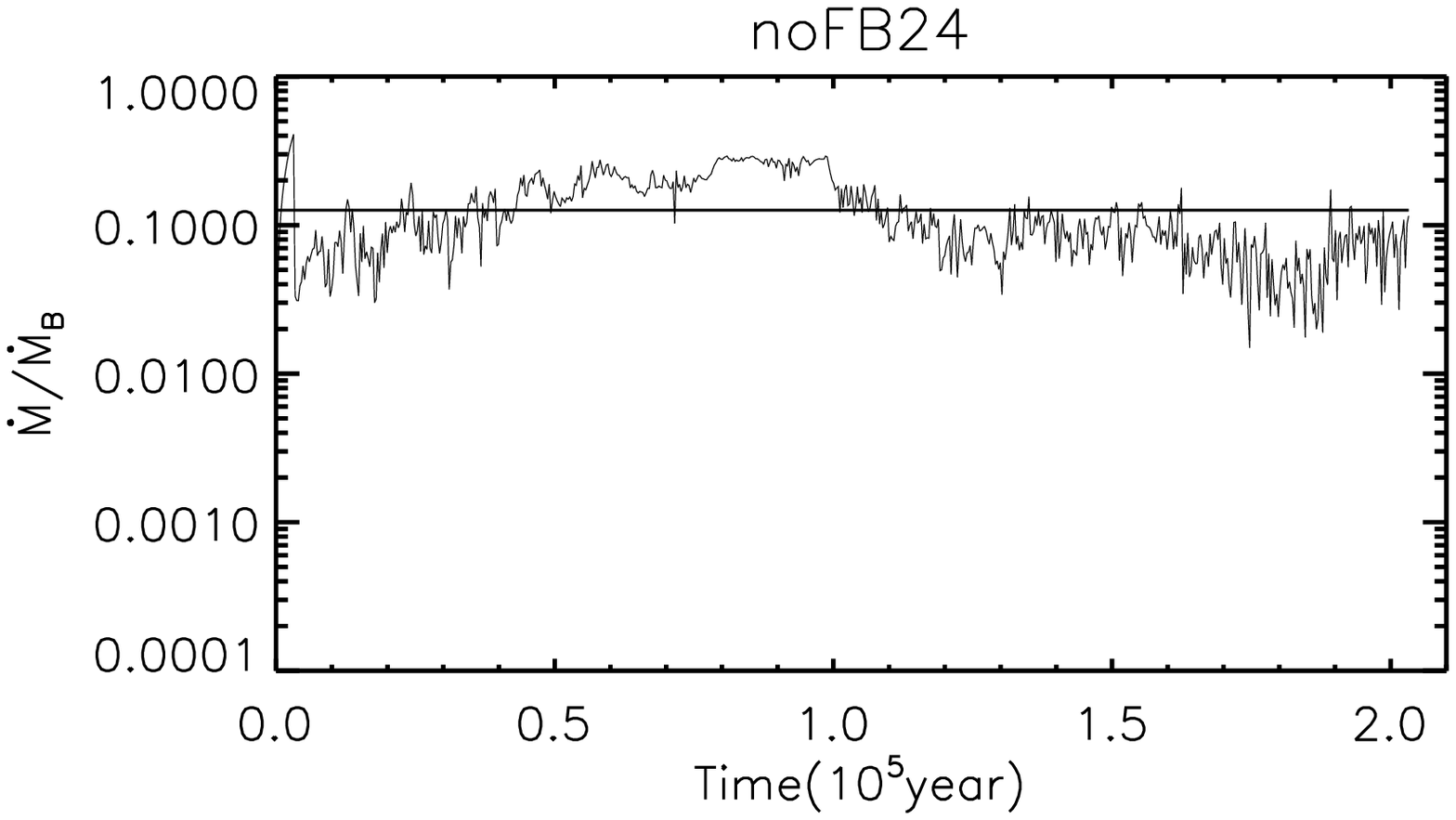}\hspace*{0.7cm} \\
\includegraphics[scale=0.5]{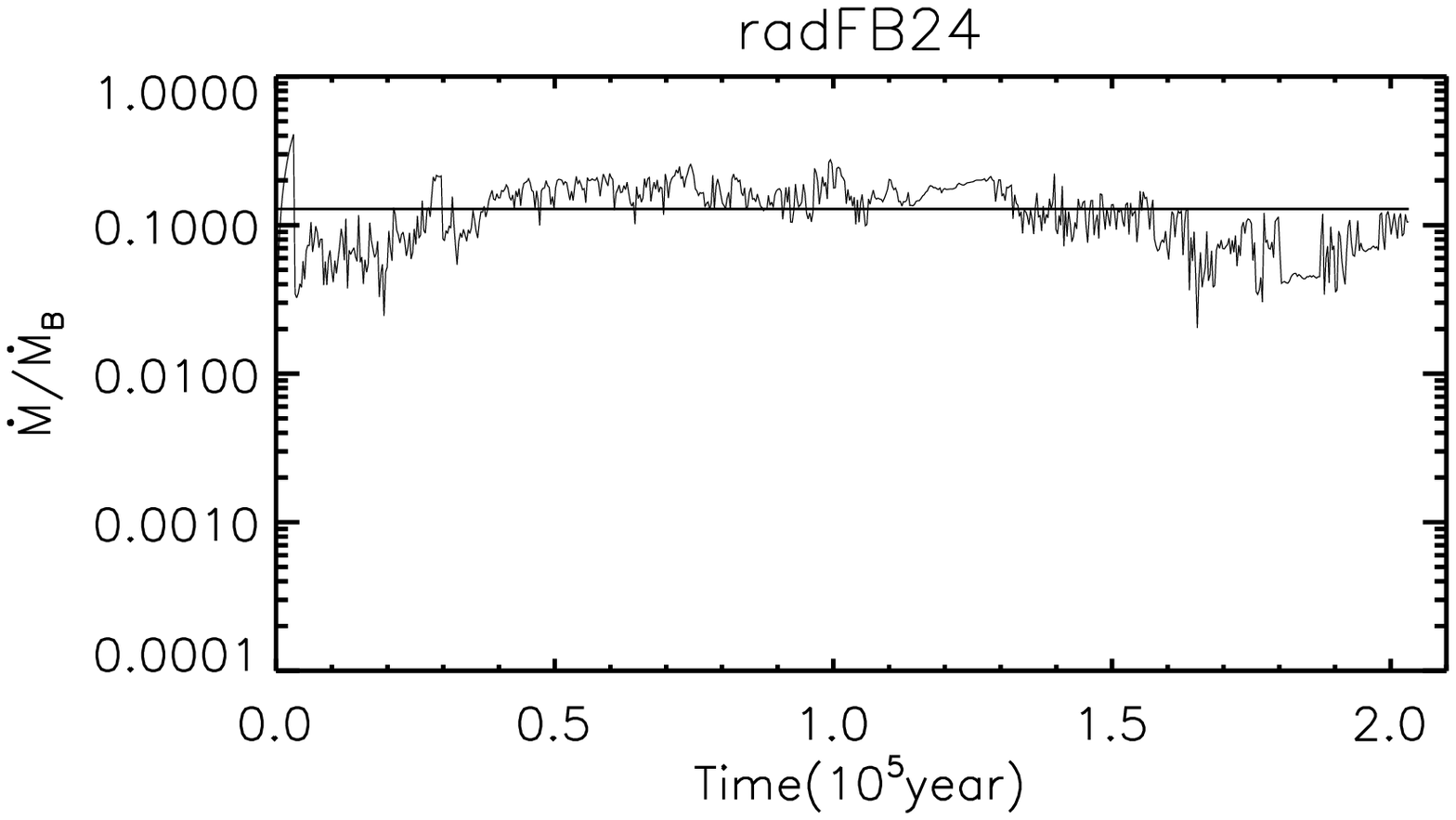}\hspace*{0.7cm} \\
\includegraphics[scale=0.5]{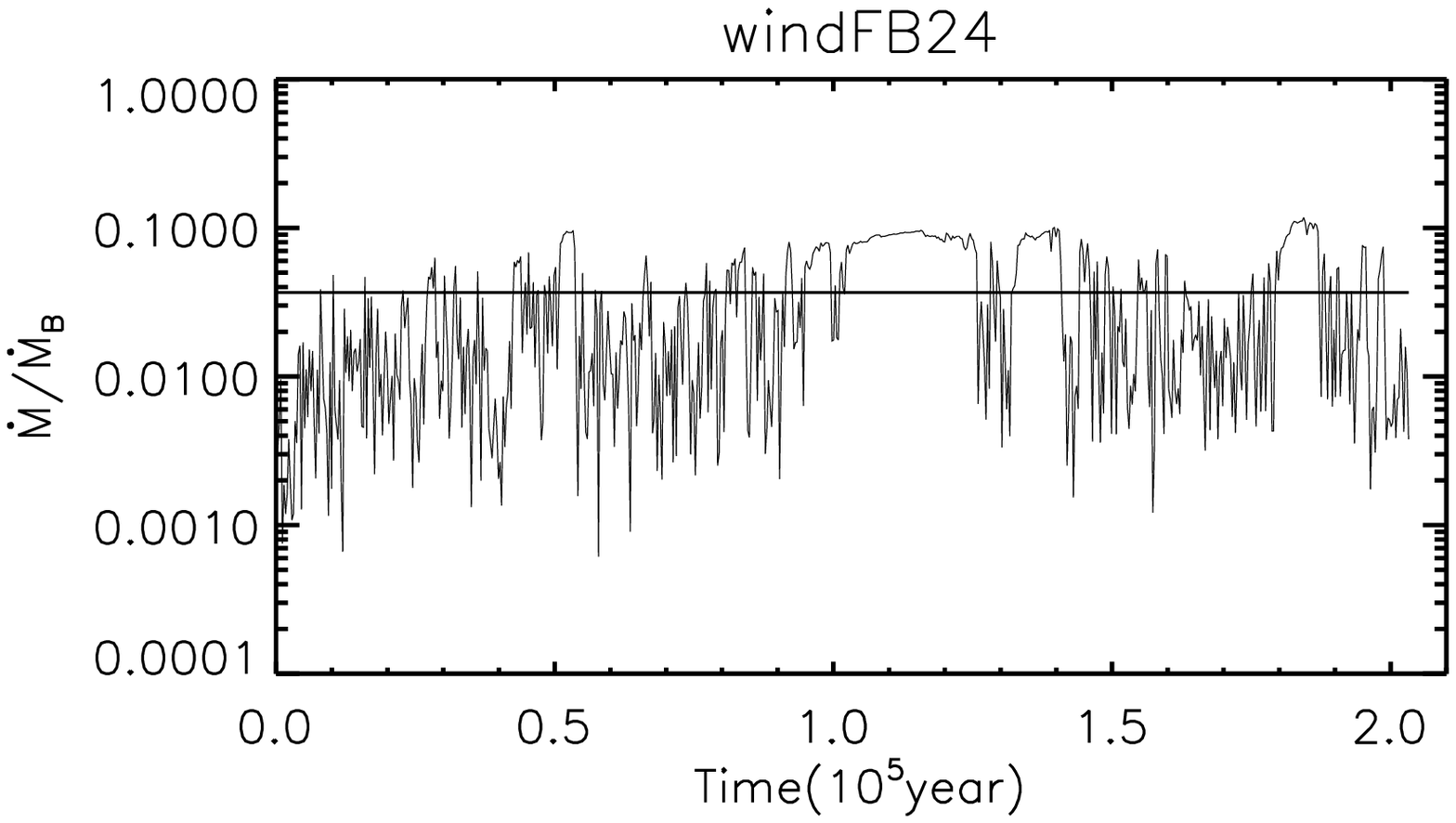}\hspace*{0.7cm} \\
\includegraphics[scale=0.5]{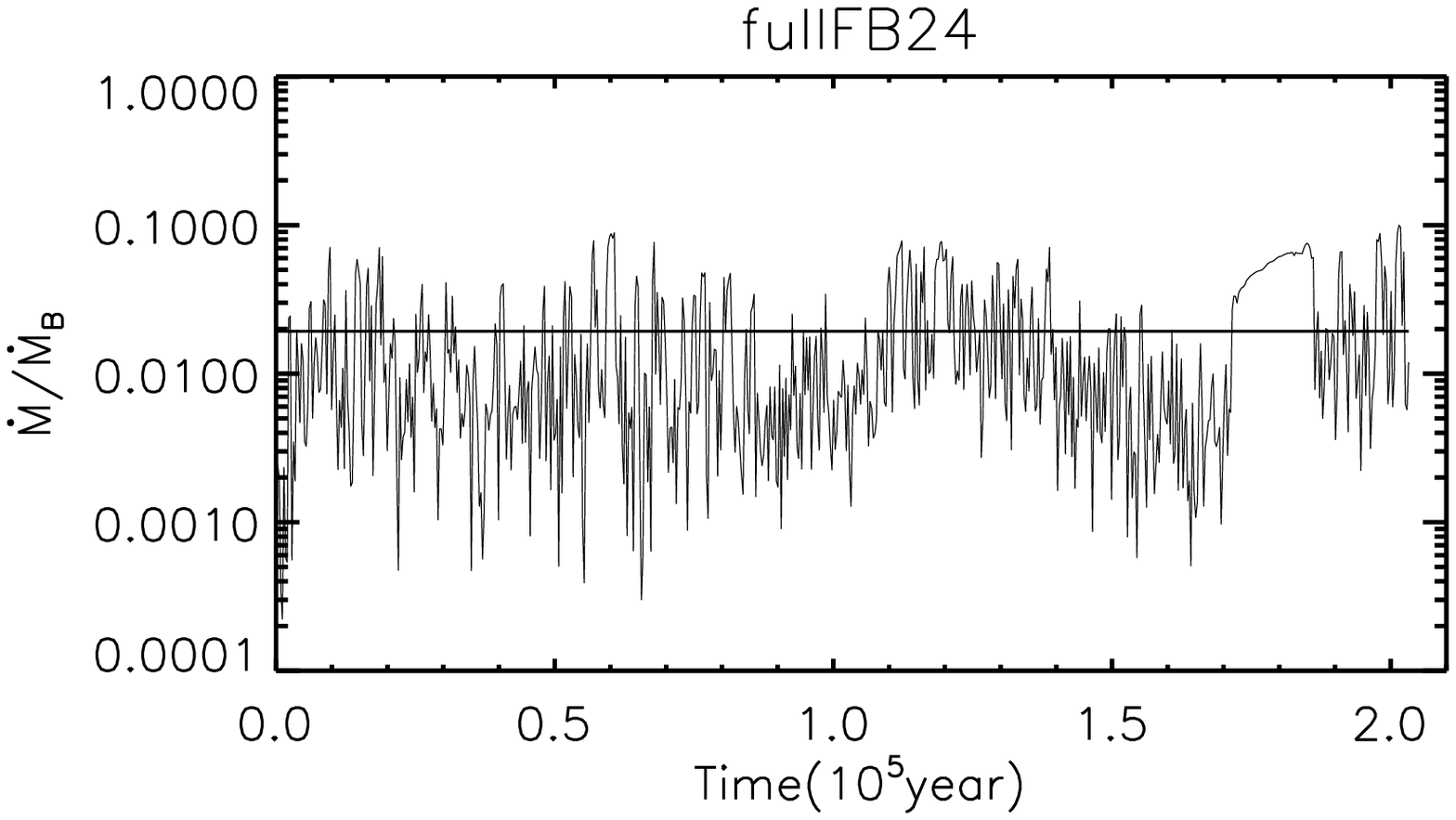}\hspace*{0.7cm}
\hspace*{0.5cm} \caption{Time evolution of mass accretion rate (in unit of Bondi accretion rate $\dot M_{\rm B}$) measured at the inner boundary. Bondi accretion rate is calculated by setting $\gamma=5/3$. The horizontal solid line corresponds to the time-averaged (from $t=0$ to $2 \times 10^5$ year) value. From top to bottom, the panels correspond to models noFB24, radFB24, windFB24 and fullFB24, respectively.  \label{Fig:mdot24}}
\end{center}
\end{figure}

For models with $\rho_0=10^{-24}{\rm g/cm^3}$, the accretion rate is significantly smaller than $\dot M_{\rm Edd}$. Bremsstrahlung radiative cooling is not important. When we calculate the Bondi accretion rate, we set $\gamma=5/3$. Figure \ref{Fig:mdot24} shows the time evolution of mass mass accretion rate (in unit of $\dot M_{\rm B}$) measured at the inner boundary. The horizontal solid line in each panel corresponds to the time-averaged value. From top to bottom, the panels correspond to models noFB24, radFB24, windFB24 and fullFB24, respectively.

For the no feedback model (noFB24), one would expect the mass accretion rate should equal to the Bondi accretion rate. However, we find that in this model, outflow is present. Due to the presence of outflow, the mass accretion rate is much smaller than the Bondi accretion rate. In our models, we inject gas with $T=7\times 10^6K$ at the radial outer boundary. For $T=7\times 10^6K$, the Bondi radius is located at $4.6 \times 10^5r_s$. Therefore, the injected gas at the outer boundary ($10^6r_s$) has positive Bernoulli parameter. The mass accretion rate in this model is very low. The bremsstrahlung cooling timescale is much longer than gas infall timescale ($r/v_r$). Therefore, radiative cooling can be neglected. The gas has enough energy to form outflows. Proga \& Begelman (2003) studied low angular momentum hot accretion flow without radiative cooling, they also find that outflow can be produced.

\begin{figure}
\begin{center}
\includegraphics[scale=0.5]{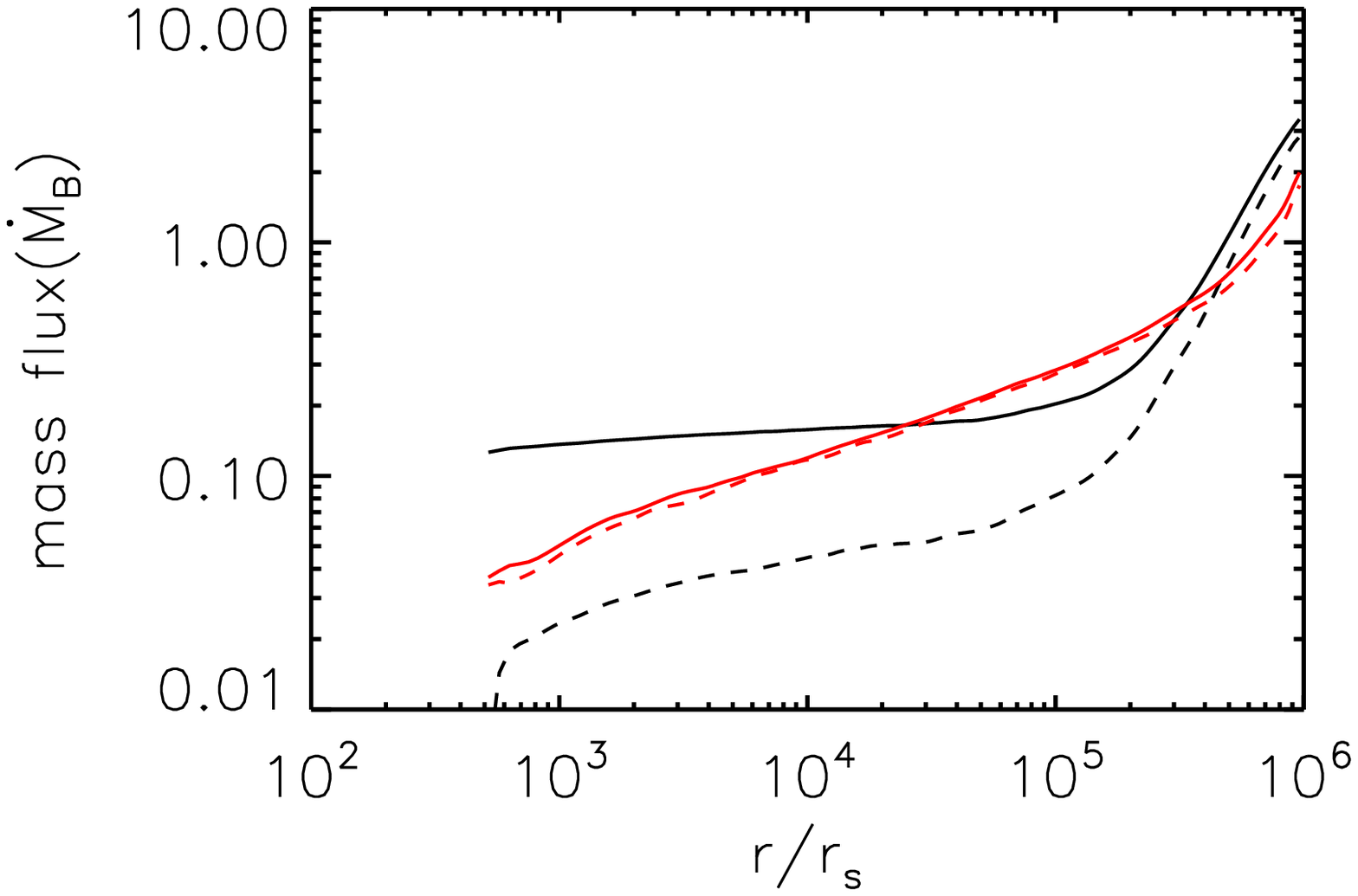}\hspace*{0.7cm} \\
\includegraphics[scale=0.5]{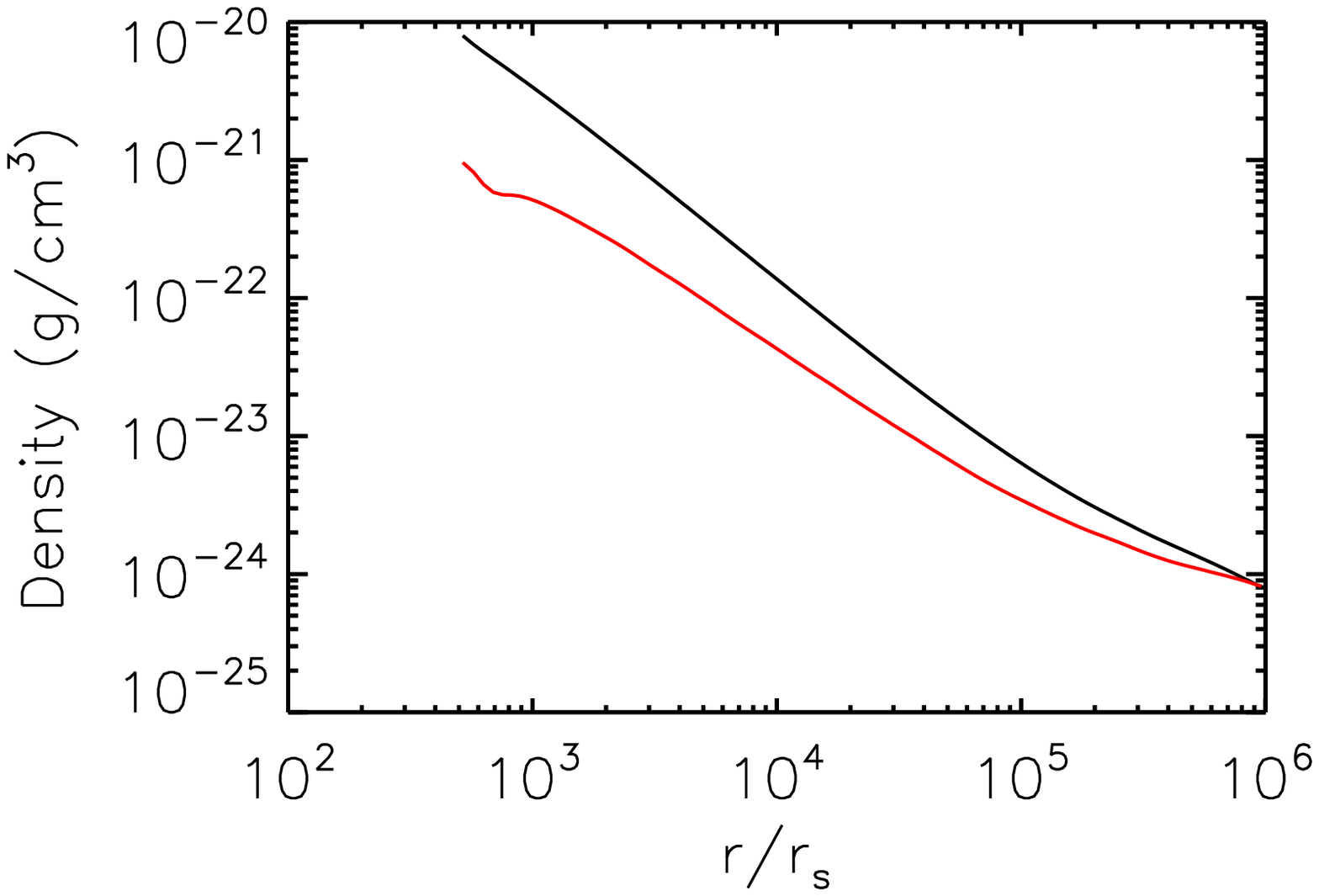}\hspace*{0.7cm} \\
\includegraphics[scale=0.5]{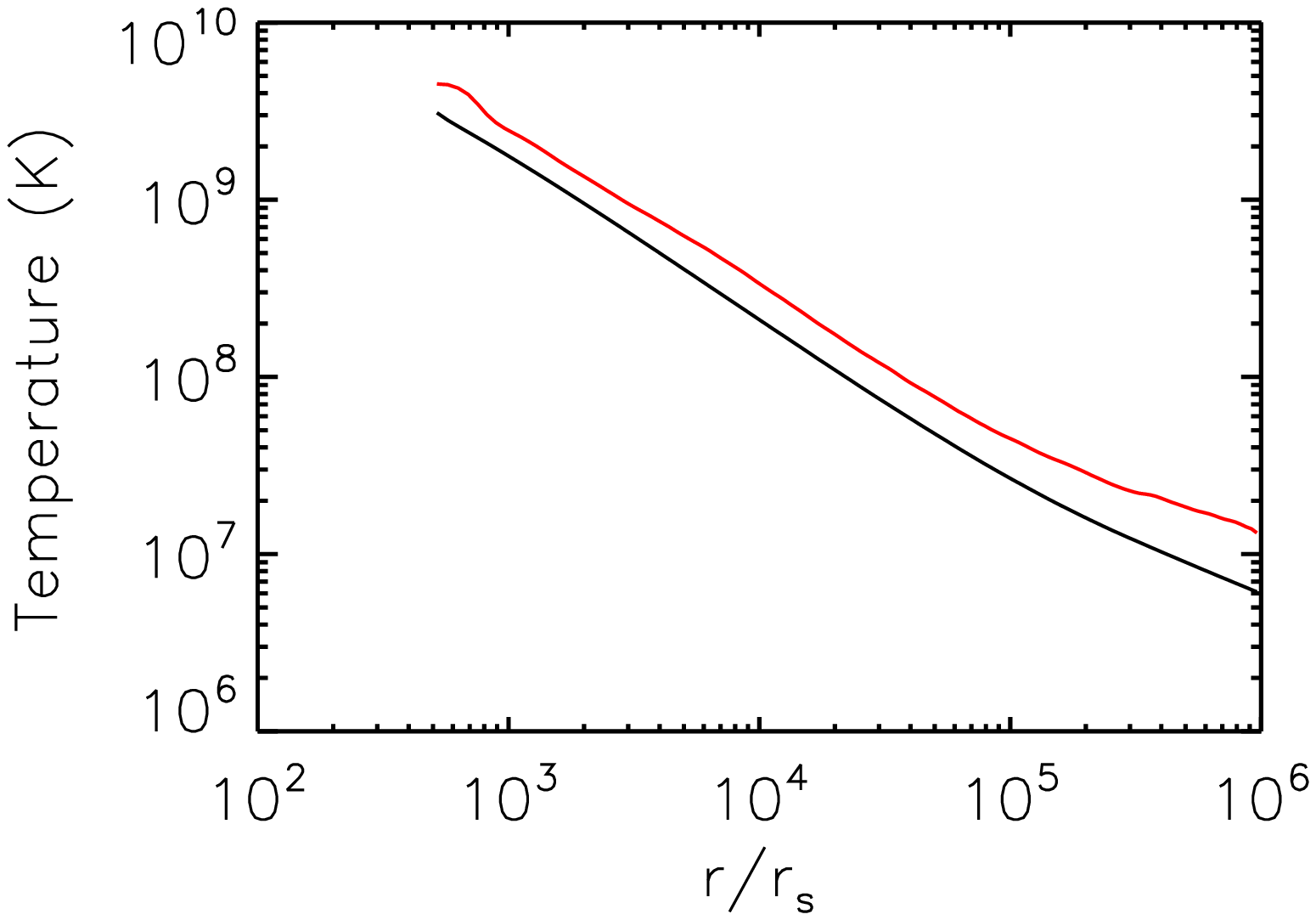}\hspace*{0.7cm}
\hspace*{0.5cm} \caption{Top panel: radial profiles of time-averaged (from $t=0$ to $2 \times 10^5$ year) mass inflow rate (solid lines) and outflow rate (dashed lines) for models noFB24 (black lines) and windFB24 (red lines). The mass fluxes are in unit of Bondi accretion rate ($\dot M_{\rm B}$). Middle panel: radial profiles of time (from $t=0$ to $2 \times 10^5$ year) and $\theta$ (from $\theta=0^\circ$ to $90^\circ$) averaged gas density for models noFB24 (black line) and windFB24 (red line). Bottom panel: radial profiles of time (from $t=0$ to $2 \times 10^5$ year) and $\theta$ (from $\theta=0^\circ$ to $90^\circ$) averaged gas temperature for models noFB24 (black line) and windFB24 (red line). \label{Fig:density24}}
\end{center}
\end{figure}

\begin{figure*}
\begin{center}
\includegraphics[scale=0.5]{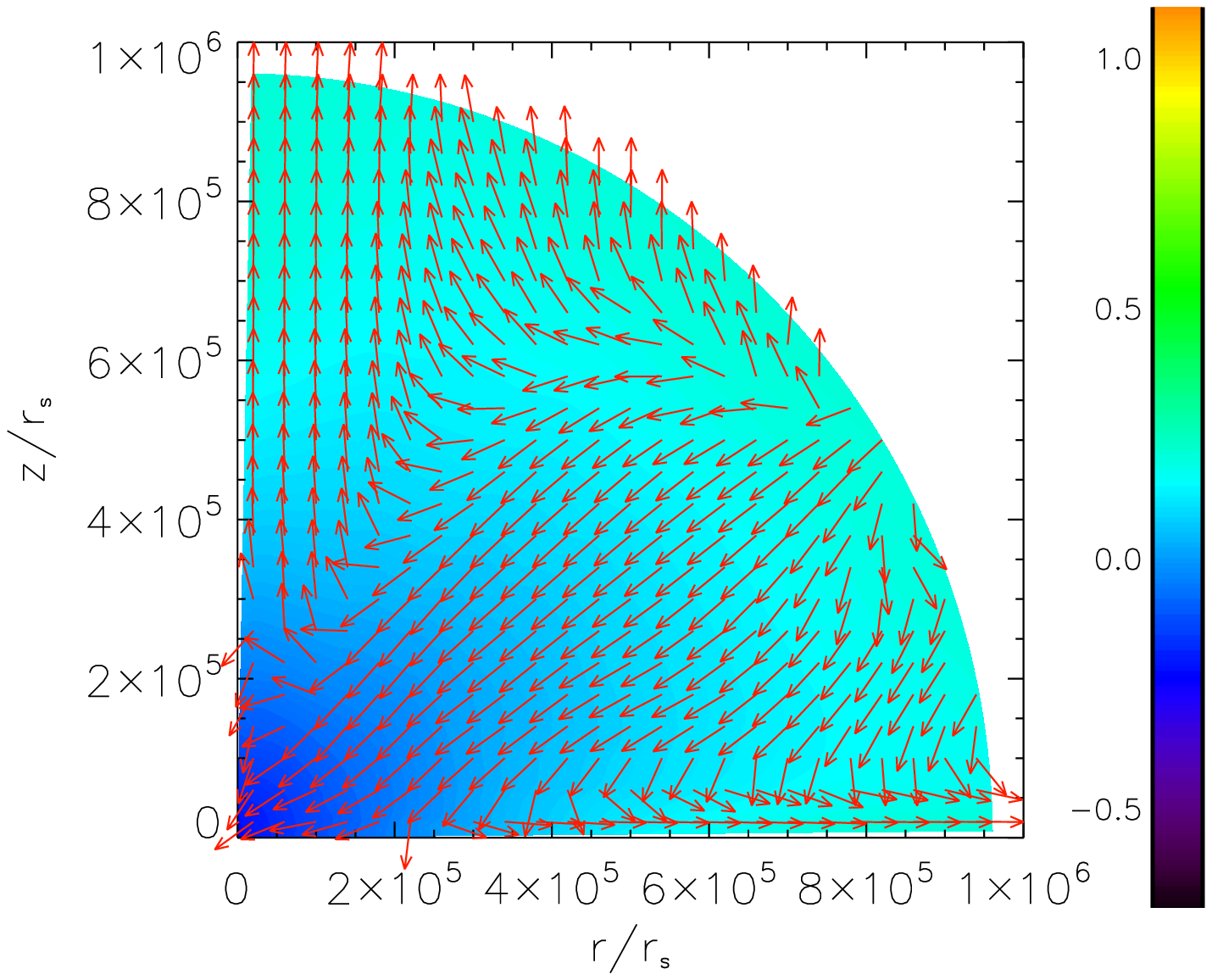}\hspace*{0.7cm}
\includegraphics[scale=0.5]{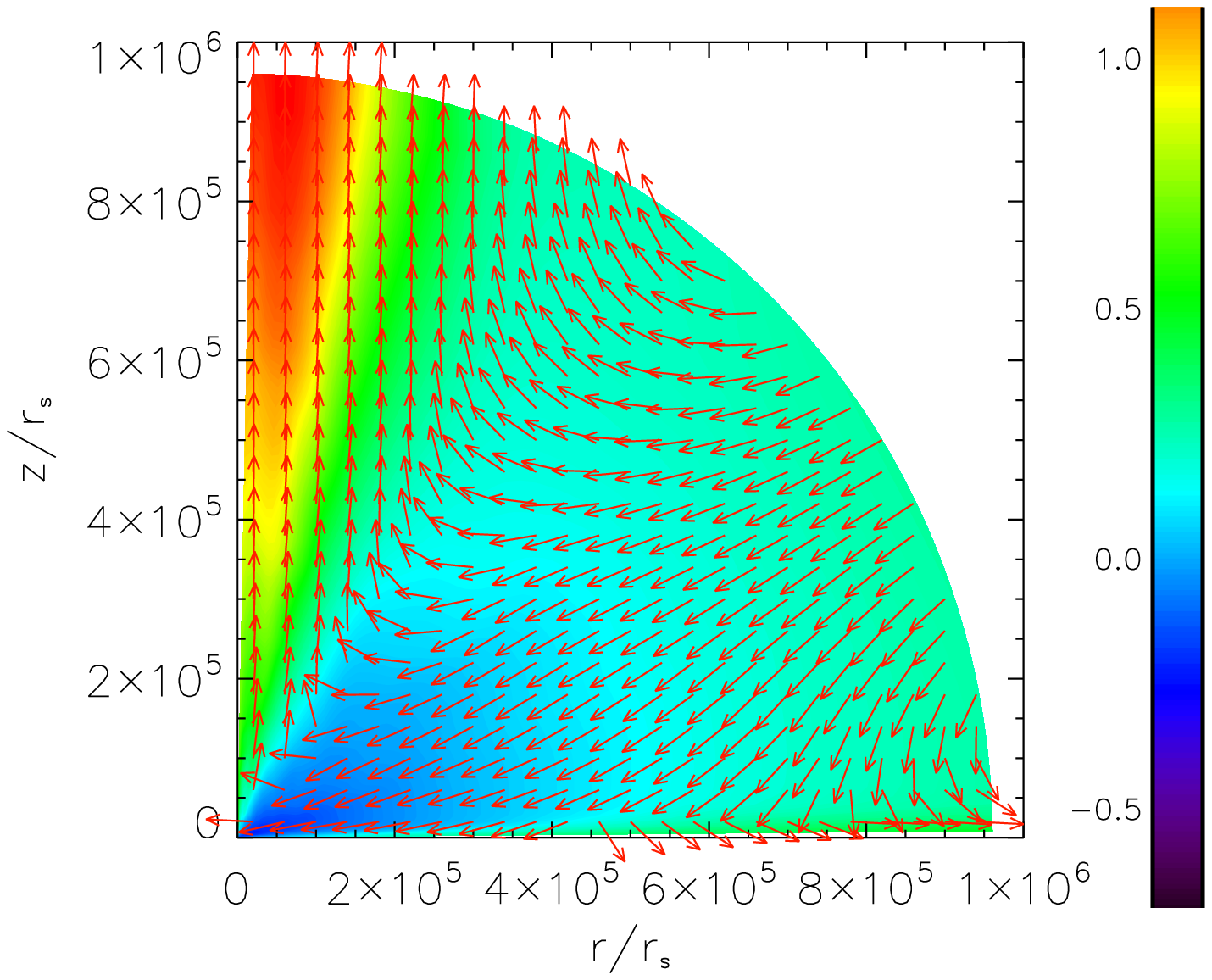}\hspace*{0.7cm}
\hspace*{0.5cm} \caption{Two-dimensional time-averaged (from $t=0$ to $2 \times 10^5$ year) properties of models noFB24 (left panel) and windFB24 (right panel). Colors show logarithm temperature in unit of virial temperature.; vectors show unit velocity vector.  \label{Fig:vector24}}
\end{center}
\end{figure*}

\begin{figure}
\begin{center}
\includegraphics[scale=0.5]{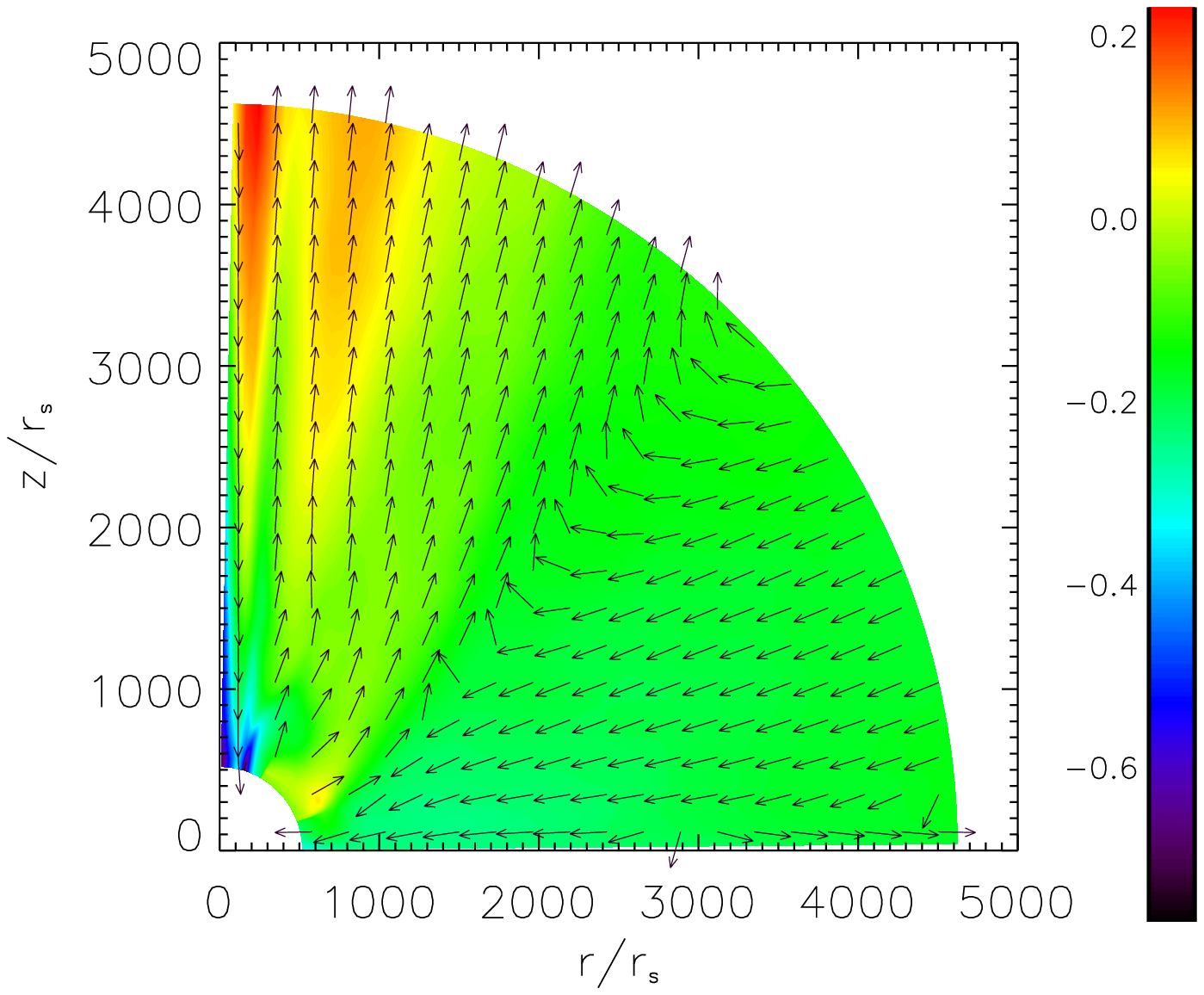}\hspace*{0.7cm}
\hspace*{0.5cm} \caption{A zooming in of right panel in Figure \ref{Fig:vector24}. \label{Fig:vectorzoomwindFB24}}
\end{center}
\end{figure}

Comparing the models noFB24 and radFB24 shown in Figure \ref{Fig:mdot24}, we see that the evolution patterns of accretion rate are roughly same. Also, both the magnitude of fluctuation and the time-averaged values of mass accretion rate are also same. Therefore, radiation feedback plays a negligible role. The reasons are as follows. For the radiation feedback, we consider both the radiation pressure due to Thompson scattering and Compton heating/cooling. The luminosity in model radFB24 is significantly smaller than Eddington value (see Table 1). Therefore, the radiation pressure is negligibly small compared to gravity. We have calculated the time and $\theta$ averaged value for density in model radFB24, we find that the averaged density decreases with increasing radius. The averaged value of density is in the range $10^{-24} {\rm g/cm^3}-10^{-20} {\rm g/cm^3}$. According to Equation (16), the length scale of radiation feedback is $\sim 10^7-10^{11} r_s$, which is significantly larger than the outer boundary of our simulation. Radiation can not effectively deposit its energy to the accretion flow. We have also calculated the optical depth of Thompson scattering using the time-averaged density, we find that the scattering optical depth is $\tau \sim 1.4 \times 10^{-4}$. Therefore, radiation can only deposit $\sim 10^{-4}$ of its energy in the accretion flow. We have also compared the Compton heating timescale ($e/S_c$) to the gas infall timescale ($r/v_r$). We find that the Compton heating timescale is several orders of magnitude longer than the gas infall timescale. Therefore, radiation feedback is not important.

In model windFB24, the time-averaged $\rho_{-24}^{-1}T_7^2$ is in the range $1-100$.  According to Equation (17), the length scale of wind feedback is $\sim 5\times 10^{5}-5 \times 10^{7} r_s$. It is much smaller than that of radiation feedback. Wind can more efficiently interact with the accretion flow. From Figure \ref{Fig:mdot24}, we see that the time-averaged mass accretion rate is $3.2$ times smaller than that in model noFB24. Wind can push away the gas surrounding the LLAGN, which will decrease the mass accretion rate. In model windFB24, the maximum accretion rate can be 2 orders of magnitude higher than the minimum accretion rate. In model noFB24, the maximum accretion rate is just 1 order of magnitude higher than the minimum accretion rate. After considering wind feedback, the fluctuation magnitude of mass accretion rate is significantly enhanced. The reason for the fluctuation is as follows. When the accretion rate is high, the wind is strong. Strong wind will push the gas surrounding the LLAGN away, which will decrease the mass supply rate for the LLAGN. The LLAGN will enter into low accretion rate stage. Then, wind becomes weak, the gas will fall onto the central region again. Then, the accretion rate becomes higher again.

\begin{figure}
\begin{center}
\includegraphics[scale=0.5]{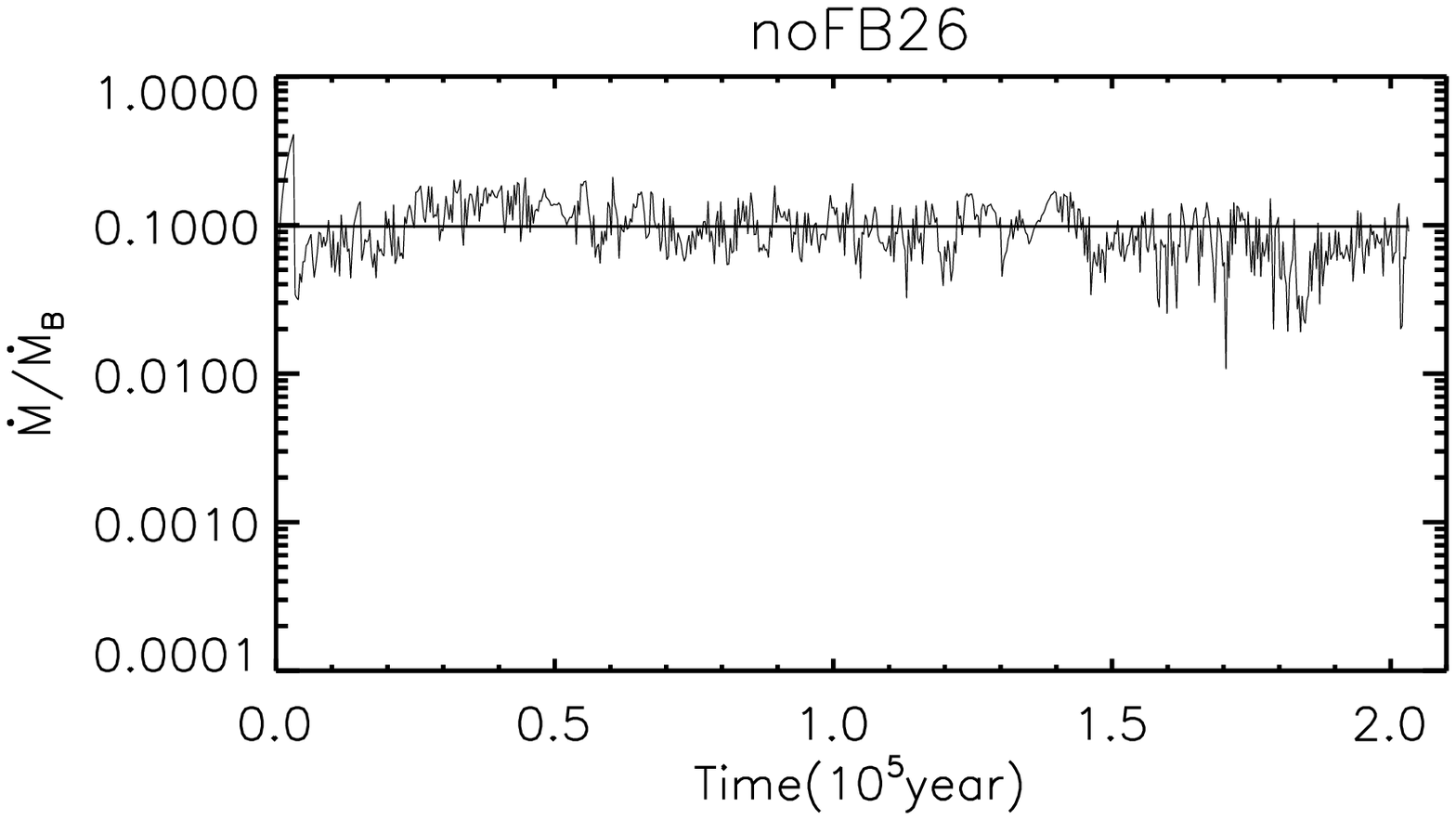}\hspace*{0.7cm} \\
\includegraphics[scale=0.5]{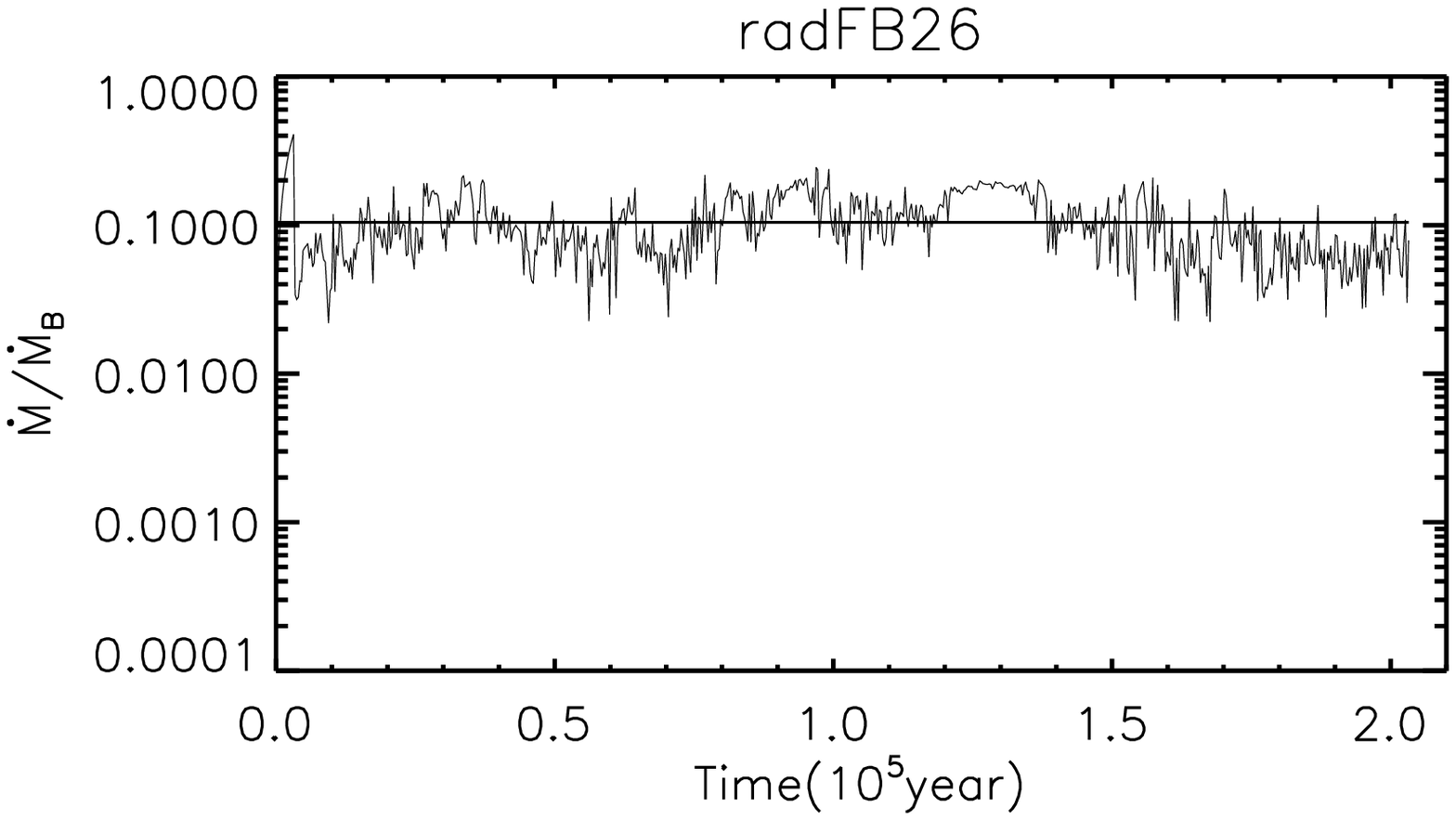}\hspace*{0.7cm} \\
\includegraphics[scale=0.5]{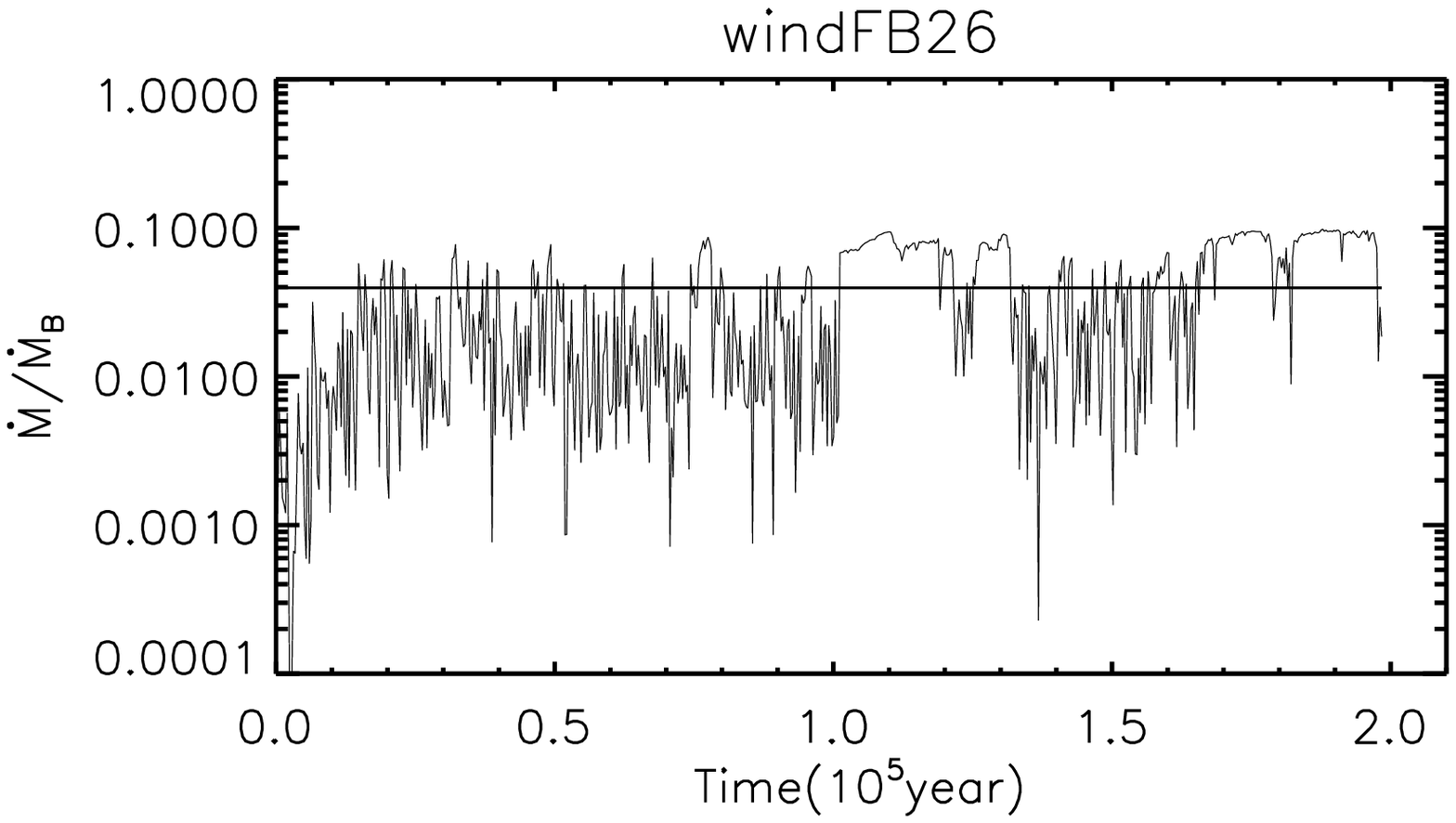}\hspace*{0.7cm} \\
\includegraphics[scale=0.5]{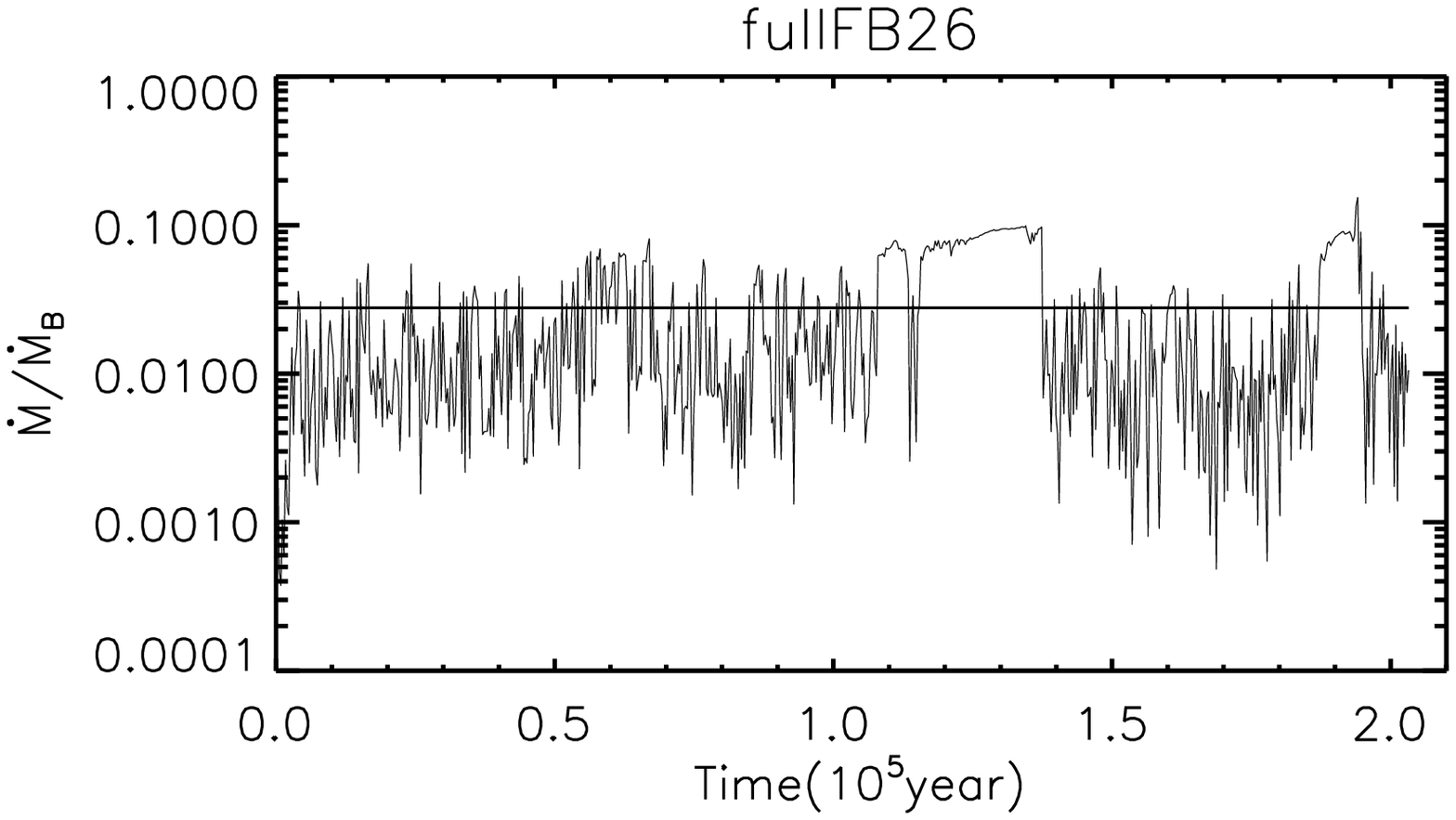}\hspace*{0.7cm}
\hspace*{0.5cm} \caption{Same as Figure \ref{Fig:mdot24}, but for models noFB26, radFB26, windFB26 and fullFB26.  \label{Fig:mdot26}}
\end{center}
\end{figure}

In order to study the detailed effects of wind feedback, we plot Figure \ref{Fig:density24}. The top panel shows radial profiles of time-averaged (from $t=0$ to $2 \times 10^5$ year) mass inflow rate (solid lines) and outflow rate (dashed lines) for models noFB24 (black lines) and windFB24 (red lines). In model noFB24, outflow is only strong outside $2\times10^5r_s$. Inside $2\times10^5r_s$, mass outflow rate is significantly smaller than mass inflow rate. In model windFB24, outflow is strong in the whole computational domain. Also, the outflow rate is almost equal to the mass inflow rate. Therefore, in model windFB24, the mass inflow rate keeps decreasing from the outer boundary to the inner boundary. The mass accretion rate at the inner boundary in model windFB24 is smaller than that in model noFB24. Due to stronger outflow, gas density in model windFB24 is much smaller than that in model noFB24 (see middle panel of Figure \ref{Fig:density24}). The wind feedback can also convert some energy of wind to the internal energy of accretion flow. Therefore, gas temperature in model windFB24 is higher than that in model noFB24 (see bottom panel of Figure \ref{Fig:density24}).

In order to see the geometry of outflow, we plot Figures \ref{Fig:vector24} and \ref{Fig:vectorzoomwindFB24}. In Figure \ref{Fig:vector24}, we plot two-dimensional time-averaged (from $t=0$ to $2 \times 10^5$ year) properties of models noFB24 (left panel) and windFB24 (right panel). Colors show logarithm temperature in unit of virial temperature; vectors show unit velocity vector. Figure \ref{Fig:vectorzoomwindFB24} is a zooming in of the right panel of Figure \ref{Fig:vector24}. From these figures, we see that the time-averaged flow is very ordered. However, we note that the snapshot of the accretion flows is very disordered. For the model noFB24, outflow is clearly present outside $2\times 10^5r_s$, which is consistent with that introduced above (see top panel of Figure \ref{Fig:density24}). From the right panel of Figure \ref{Fig:vector24} and Figure \ref{Fig:vectorzoomwindFB24}, we see that in model windFB24, outflow is present from the inner boundary to the outer boundary. Wind is injected in the region $30^\circ<\theta<70^\circ$. When it moves outward, it changes its direction to $0^\circ<\theta<45^\circ$. When wind moves outwards, more and more gas from inflowing region joins wind. Therefore, the outflow mass flux increases with increasing radius.

\begin{figure}
\begin{center}
\includegraphics[scale=0.5]{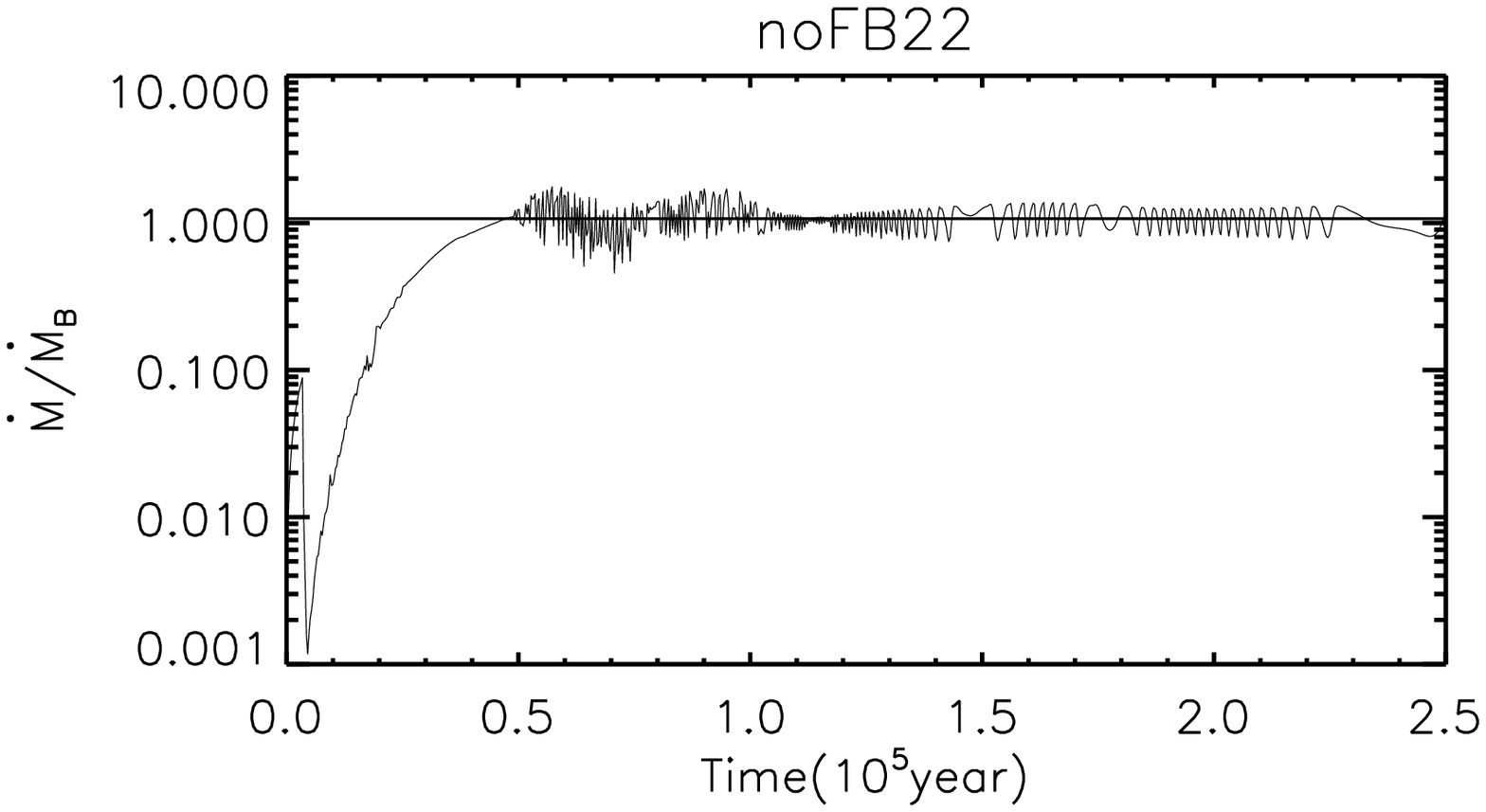}\hspace*{0.7cm} \\
\includegraphics[scale=0.5]{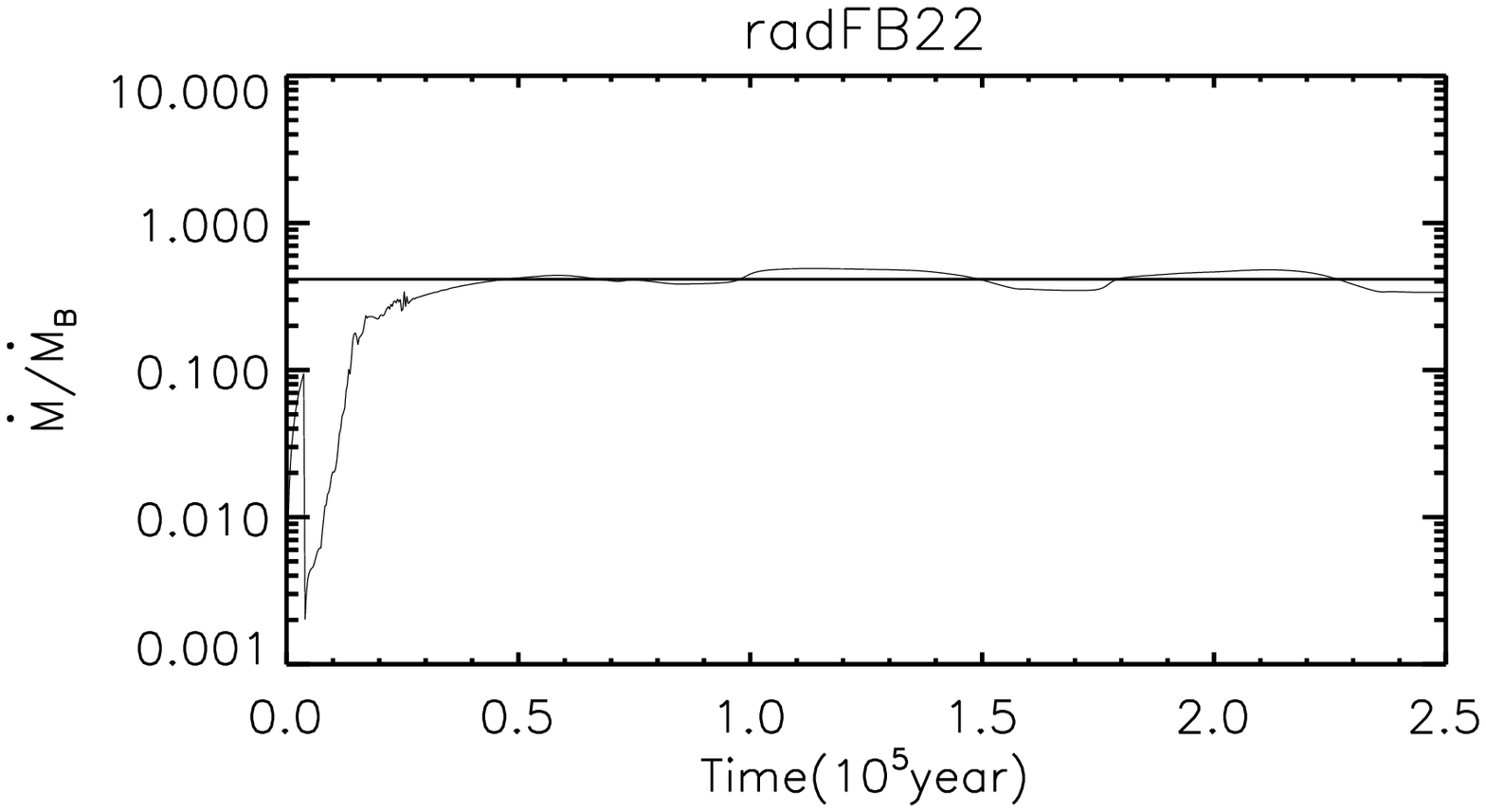}\hspace*{0.7cm} \\
\includegraphics[scale=0.5]{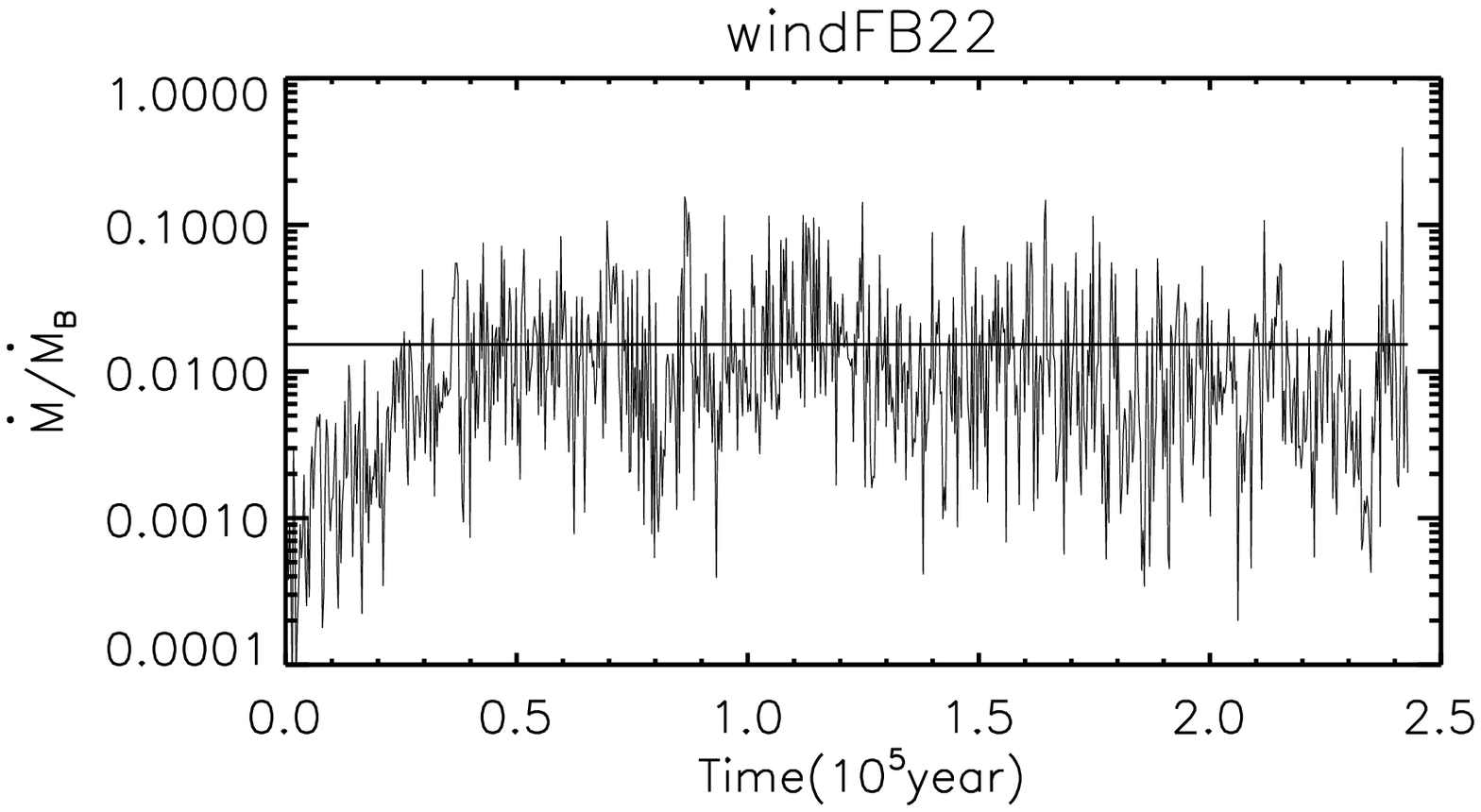}\hspace*{0.7cm} \\
\includegraphics[scale=0.5]{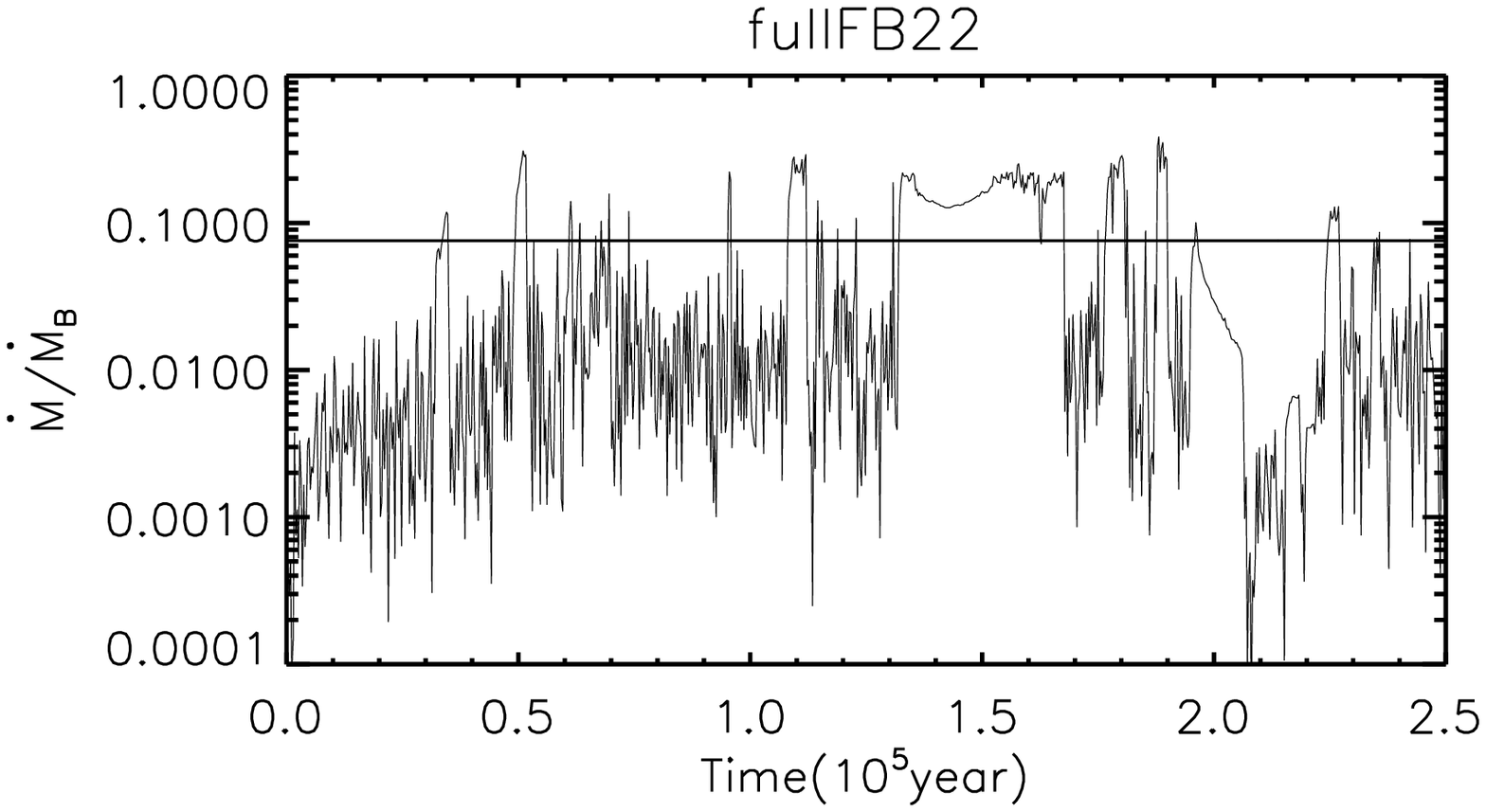}\hspace*{0.7cm}
\hspace*{0.5cm} \caption{Time evolution of mass accretion rate (in unit of Bondi accretion rate $\dot M_{\rm B}$) measured at the inner boundary. Bondi accretion rate is calculated by setting $\gamma=1$. The horizontal solid line corresponds to the time-averaged (from $t=1.2\times 10^5$ to $2.5 \times 10^5$ year) value. From top to bottom, the panels correspond to models noFB22, radFB22, windFB22 and fullFB22, respectively.  \label{Fig:mdot22}}
\end{center}
\end{figure}

\begin{figure}
\begin{center}
\includegraphics[scale=0.5]{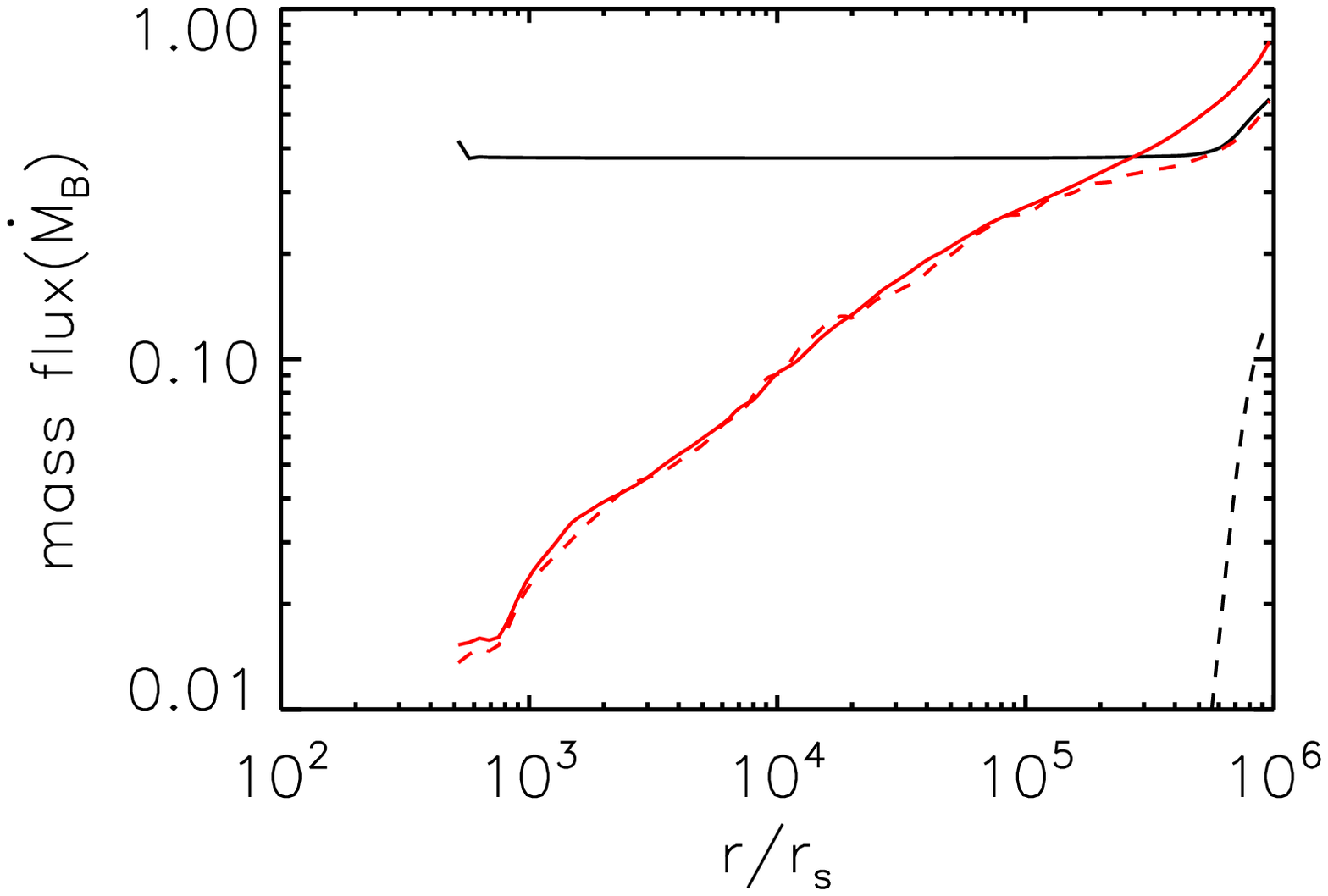}\hspace*{0.7cm} \\
\includegraphics[scale=0.5]{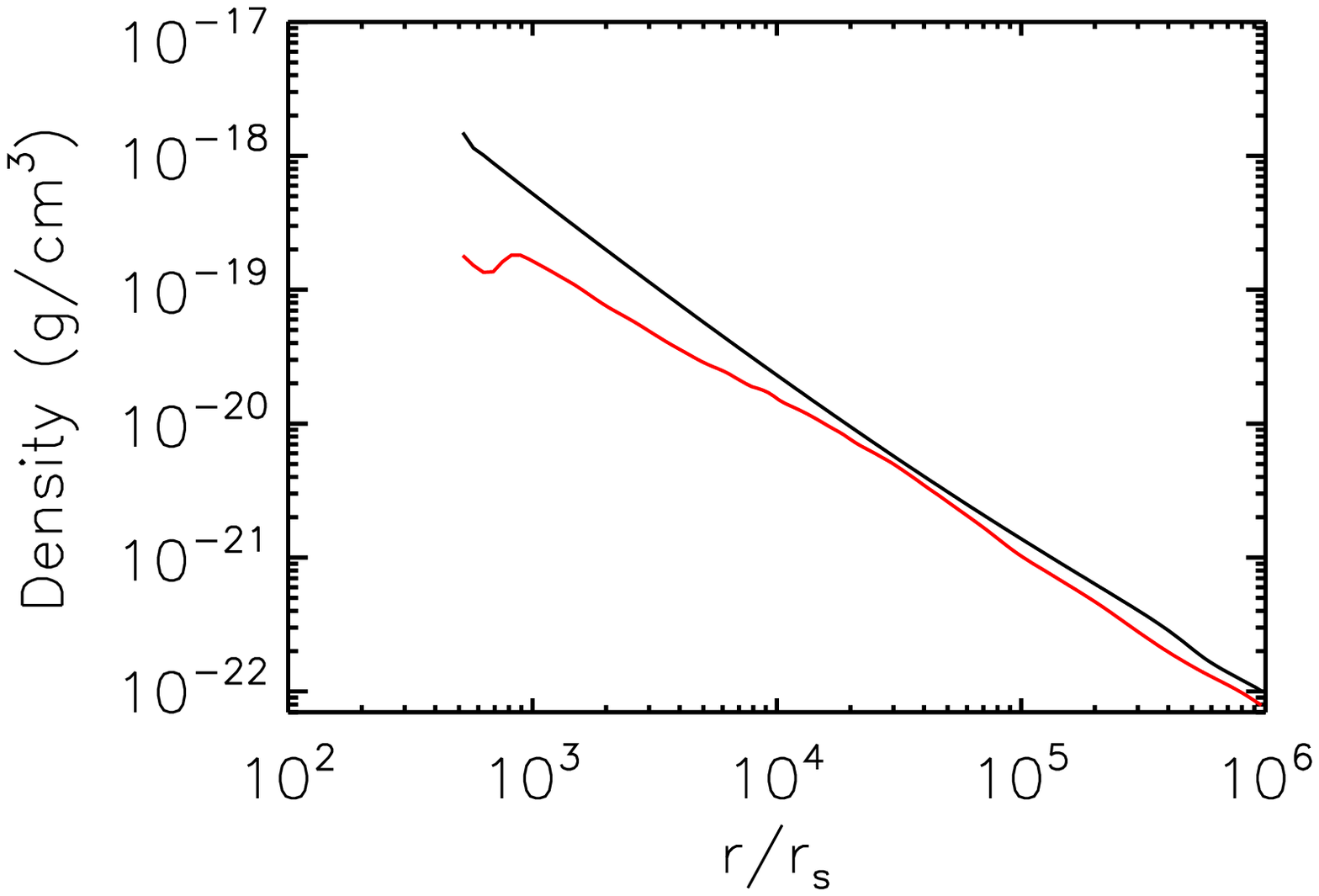}\hspace*{0.7cm} \\
\includegraphics[scale=0.5]{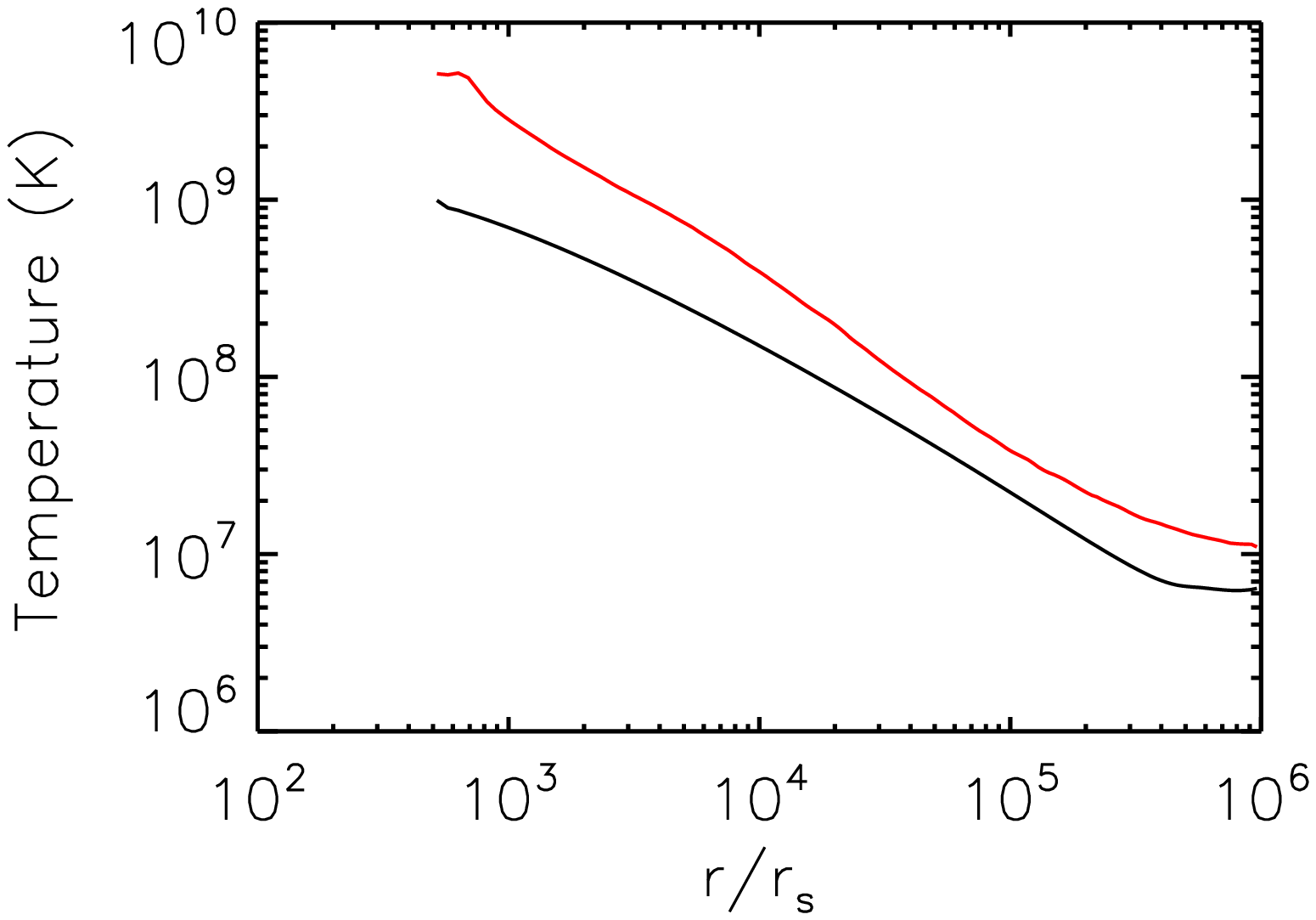}\hspace*{0.7cm}
\hspace*{0.5cm} \caption{Top panel: radial profiles of time-averaged (from $t=1.2$ to $2.5 \times 10^5$ year) mass inflow rate (solid lines) and outflow rate (dashed lines) for models radFB22 (black lines) and windFB22 (red lines). The mass fluxes are in unit of Bondi accretion rate ($\dot M_{\rm B}$). Middle panel: radial profiles of time (from $t=1.2$ to $2.5 \times 10^5$ year) and $\theta$ (from $\theta=0^\circ$ to $90^\circ$) averaged gas density for models radFB22 (black line) and windFB22 (red line). Bottom panel: radial profiles of time (from $t=1.2$ to $2.5 \times 10^5$ year) and $\theta$ (from $\theta=0^\circ$ to $90^\circ$) averaged gas temperature for models radFB22 (black line) and windFB22 (red line). \label{Fig:density22}}
\end{center}
\end{figure}

Comparing models windFB24 and fullFB24 (see Figure \ref{Fig:mdot24}), we can see that the magnitude of mass accretion rate fluctuation in these two models is roughly same. This indicates that the fluctuation is mainly due to the effect of wind feedback. The averaged mass accretion rate in model fullFB24 is $\sim 2$ times smaller than that in model windFB24. Radiation plays some role in reducing the mass accretion rate in model fullFB24. We note that the two-dimensional geometry of outflow in model fullFB24 is quite similar as that in model windFB24 shown in Figure \ref{Fig:vector24}. Therefore, we do not show it here.

The black hole luminosity listed in Table 2 is calculated using the radiative efficiency shown in Equation (11). The radiative efficiency decreases with the decreasing of mass accretion rate. Comparing model noFB24 to model fullFB24, we see that when feedback is taken into account, the mass accretion rate can be decreased by a factor $\sim 6.4$. However, with the decrease of accretion rate, radiative efficiency is also decreased. Therefore, the black hole luminosity in model fullFB24 is $16$ times smaller than that in model noFB24. If the black hole accretes at the Bondi rate $\dot M(r_{\rm in})=\dot M_B$, we will have black hole luminosity $L=10^{-5}L_{\rm Edd}$. In the model without feedback (noFB24), due to the presence of outflow, the accretion rate is much smaller than the Bondi value. Correspondingly, the luminosity is decreased by a factor of $25$. When feedback is considered, the luminosity is decreased by another factor of $\sim 16$. The black hole luminosity in model fullFB24 is $\sim 400$ times smaller than that predicted by Bondi formula.

Mechanical energy flux of outflow is an important parameter in AGN feedback study. The mechanical energy flux of outflow is calculated as follows,
\begin{equation}
P_{\rm K} (r)=2\pi r^2 \int_{\rm 0^\circ}^{\rm 90^\circ} \rho
\max(v_r^3,0) \sin\theta d\theta
\end{equation}
The mechanical energy flux of outflow measured at outer boundary for all of our models is summarized in column 7 of Table 2.

We also perform simulations with lower density ($\rho_0=10^{-26}{\rm g/cm^3}$), they are models of noFB26, radFB26, windFB26 and fullFB26. We find that the results are quite similar to those in models with $\rho_0=10^{-24}{\rm g/cm^3}$. The reason is as follows. In all models with $\rho_0=10^{-26}{\rm g/cm^3}$ and $\rho_0=10^{-24}{\rm g/cm^3}$, the mass accretion rate is significantly smaller than the Eddington rate. Therefore, in these models, the bremsstrahlung radiative cooling rate is all negligibly small. As mentioned above, radiation feedback is not important in models with $\rho = 10^{-24}g/cm^3$. According to Equation (16), the length scale of radiation feedback in models radFB26 and fullFB26 is even longer than that in models radFB24 and fullFB24. Therefore, radiation feedback in all models with $\rho_0=10^{-26}{\rm g/cm^3}$ and $\rho_0=10^{-24}{\rm g/cm^3}$ is also not important. The accretion flows in these models are all evolving adiabatically. Therefore, from Figures \ref{Fig:mdot24} and \ref{Fig:mdot26}, we see that time averaged value of accretion rate (in unit of Bondi rate) and magnitude of accretion rate fluctuations are all quite similar in models noFB24, radFB24, noFB26, radFB26. The results of model windFB26 are quite similar as those of model fullFB26. Other properties of accretion flow (e.g., velocity field, density and temperature profiles) in models with $\rho_0=10^{-26}{\rm g/cm^3}$ are all quite similar as those in their counterpart models with $\rho_0=10^{-24}{\rm g/cm^3}$.

\subsection{Models with $\rho_0=10^{-22}{\rm g/cm^3}$}

If radiative cooling is not important, when gas falls inwards, gas temperature will increase. For two models with $\rho_0=10^{-22}{\rm g/cm^3}$ (noFB22, radFB22), we find that at the region around the outer boundary ($5 \times 10^5-10^6r_s$), bremsstrahlung radiative cooling timescale can be shorter than the gas infall timescale. Radiative cooling is important in this region. Gas temperature does not increase inwards. Gas temperature in this region is almost a constant with radius. Therefore, when we calculate the Bondi accretion rate in the models in this subsection, we set $\gamma=1$.

Figure \ref{Fig:mdot22} shows the time evolution of mass accretion rate. For model noFB22, the mass accretion rate quickly increases from $t=0$ to $5 \times 10^4$ year. When $t>1.2 \times 10^5$ year, the flow achieves a quasi-steady state. The mass accretion rate approximately equals to the Bondi rate. In this model, outflow does not exist. The Bernoulli parameter of the injected gas at the outer boundary is positive. When gas falls towards the center, bremsstrahlung radiation cools the gas. The Bernoulli parameter decreases inwards. We find that when $r<1.4 \times 10^5r_s$, the Bernoulli parameter becomes negative. The gas has not enough energy to form outflow. The time-averaged luminosity is $1.5\% L_{\rm Edd}$. Hot accretion flow can only be present when $L<2\%L_{\rm Edd}$ (Yuan \& Narayan 2014). We find that when $\rho_0>10^{-22}{\rm g/cm^3}$, if no feedback is considered, the luminosity of black hole will exceed $2\%L_{\rm Edd}$. Therefore, in this paper, we do not consider accretion flow with $\rho_0>10^{-22}{\rm g/cm^3}$. The reason for oscillation of mass accretion rate is as follows. In this model, the gas has rotational velocity. Also, bremsstrahlung radiative cooling is present. The flow is not in exact equilibrium. Compared with models with wind feedback, the magnitude of oscillation of mass accretion rate is very small.

For model radFB22, we also find that the mass accretion rate quickly increases from $t=0$ to $5 \times 10^4$ year. When $t> 10^5$ year, the flow achieves a quasi-steady state. We find that when $t> 10^5$ year, the mass accretion rate oscillates around its mean value. The oscillation is due to the episodic generation of outflow outside $10^5r_s$. We have introduced the properties of Compton heating launched outflow in Bu \& Yang (2018). For convenience, we also briefly introduce it here. In the region $r>10^5r_s$, the gas temperature is lower than Compton temperature of the photons emitted by the LLAGN. Therefore, gas in this region can be Compton heated. We also calculate the radiative cooling ($e/Br$) and gas infall ($r/v_r$) timescales in this region. We find that the Compton heating timescale ($e/Sc$) is shorter than both the radiative cooling and gas infall timescales . Therefore, gas temperature in the region $r> 10^5r_s$ can be Compton heated to be above local virial temperature. Outflow can form in this region. We note that the outflow is present episodically. The reason is as follows. When the accretion rate is high, the black hole luminosity is high. Therefore, the Compton heating rate is high and outflow can form. When outflow forms, gas will be taken away. The black hole accretion rate will become low. Then Compton heating will become not important due to the decrease of luminosity (or accretion rate). Outflow will disappear. The gas will cool and fall to the center again. The black hole accretion rate will become high again. Comparing this model with noFB22, we find that the time-averaged accretion rate is reduced by a factor of 2. Correspondingly, the black luminosity is also decreased by a factor of $\sim 2$.

\begin{figure*}
\begin{center}
\includegraphics[scale=0.5]{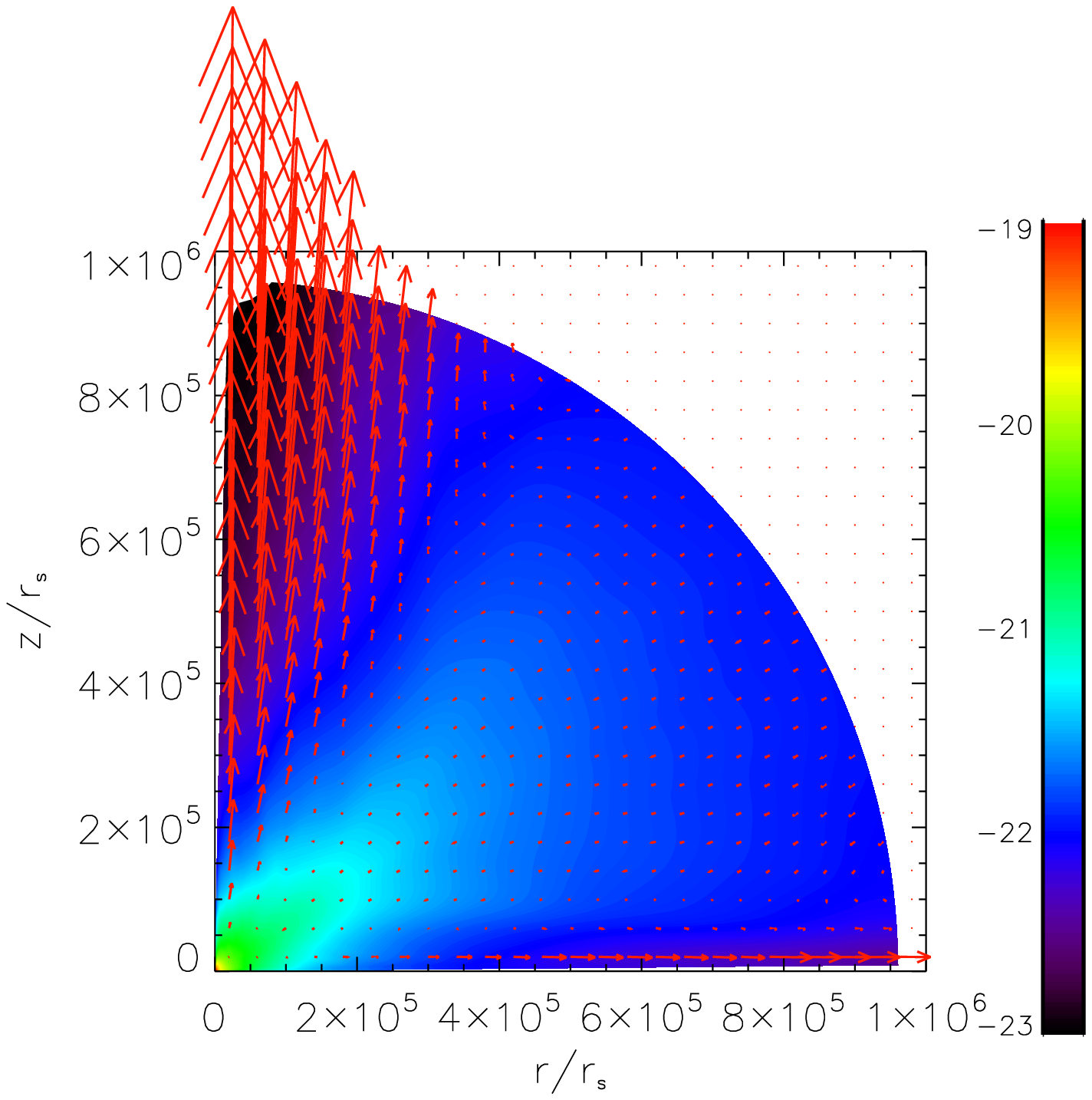}\hspace*{0.7cm}
\includegraphics[scale=0.5]{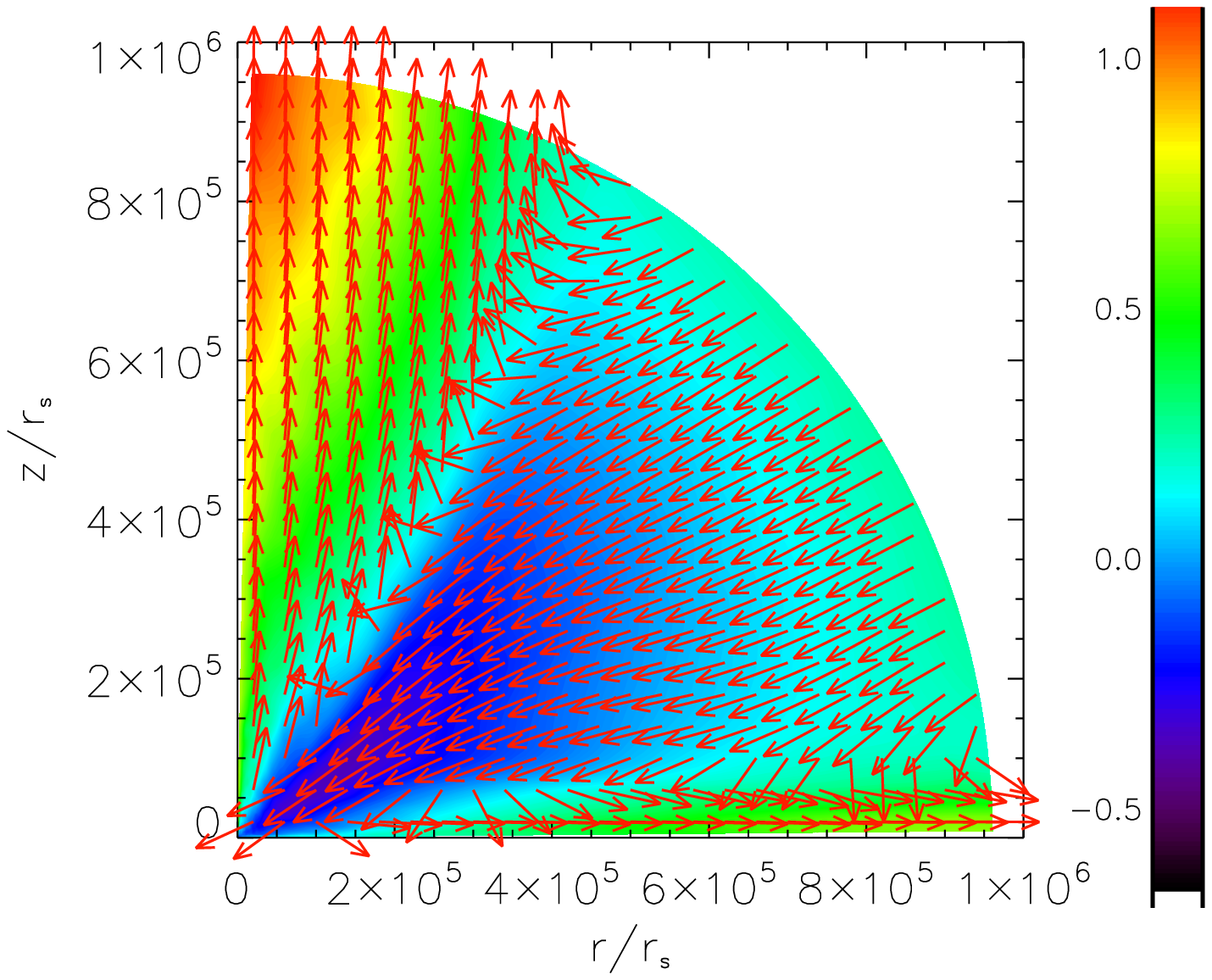}\hspace*{0.7cm}
\hspace*{0.5cm} \caption{Two-dimensional properties of model windFB22. Left panel: colors show time-averaged (from $t=1.2\times10^5$ to $2.5 \times 10^5$ year) logarithm density; vectors show time-averaged (from $t=1.2\times10^5$ to $2.5 \times 10^5$ year) velocity vector. Length of vector shows the magnitude of velocity. Right panel: colors show time-averaged (from $t=1.2\times10^5$ to $2.5 \times 10^5$ year) logarithm temperature in unit of virial temperature.; vectors show time-averaged (from $t=1.2\times10^5$ to $2.5 \times 10^5$ year) unit velocity vector.  \label{Fig:vector22}}
\end{center}
\end{figure*}

\begin{figure}
\begin{center}
\includegraphics[scale=0.5]{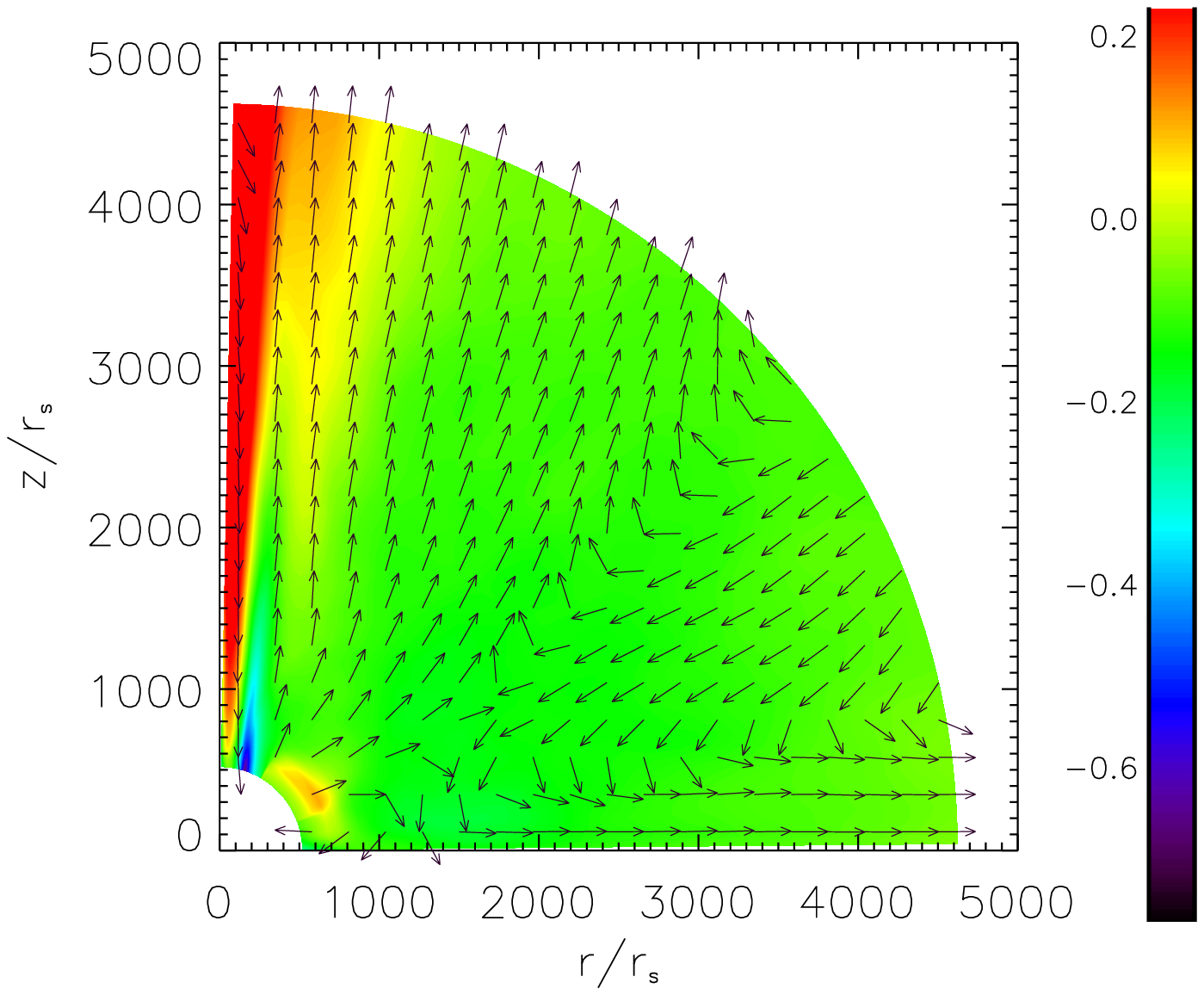}\hspace*{0.7cm}
\hspace*{0.5cm} \caption{A zooming in of right panel in Figure \ref{Fig:vector22}. \label{Fig:vectorzoomwindFB22}}
\end{center}
\end{figure}

\begin{figure}
\begin{center}
\includegraphics[scale=0.5]{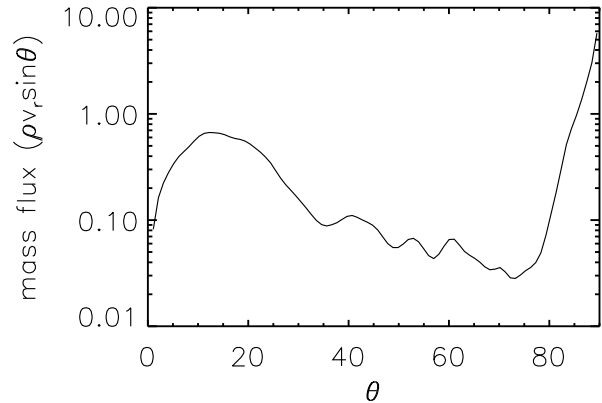}\hspace*{0.7cm}
\hspace*{0.5cm} \caption{Distribution of time-averaged (from $t=1.2\times10^5$ to $2.5 \times 10^5$ year) mass outflow rate with $\theta$ at outer radial boundary. Mass flux is in unit of Bondi accretion rate \label{Fig:windfluxtheta22}}
\end{center}
\end{figure}

Compared to the models with $\rho_0=10^{-24}{\rm g/cm^3}$, radiation feedback in models with $\rho_0=10^{-22}{\rm g/cm^3}$ is more efficient. This is because that the length scale for radiation feedback decreases with increase of gas density (Equation (16)). However, the effects of radiation feedback are still less important than those of wind feedback when $\rho_0=10^{-22}{\rm g/cm^3}$ (see next paragraph).

From the third panel of Figure \ref{Fig:mdot22} we see that in the model windFB22, wind can effectively interact with the accretion flow. Due to the presence of wind feedback, the time-averaged accretion rate is significantly decreased. The time-averaged mass accretion rate in model windFB22 is $\sim 100$ times smaller than that in model noFB22. For the models with $\rho_0=10^{-24}{\rm g/cm^3}$, wind feedback can only decrease the mass accretion rate by a factor of $\sim 3$ (see Table 1). In model windFB22, the wind power is significantly higher than that in model windFB24 (see Figure \ref{Fig:power}). Also, the length scale of wind feedback in model windFB22 is much smaller than that in model windFB24. Therefore, compared to models with $\rho_0=10^{-24}{\rm g/cm^3}$,  the wind feedback in models with $\rho_0=10^{-22}{\rm g/cm^3}$ is more efficient in suppressing the black hole accretion rate. The time-averaged black hole luminosity in model windFB22 is smaller than that in model noFB22 by a factor of $\sim 100$. In model windFB22, the reason for the fluctuation of mass accretion rate is same as that in model windFB24.

In order to study the individual effects of radiation and wind feedback, we plot Figure \ref{Fig:density22}. The top panel shows radial profiles of time-averaged (from $t=1.2 \times 10^5$ to $2.5 \times 10^5$ year) mass inflow rate (solid lines) and outflow rate (dashed lines) for models radFB22 (black lines) and windFB22 (red lines). As mentioned above, in model radFB22, outflow can be driven by Compton heating outside $10^5r_s$. Inside $10^5r_s$, Compton heating is not important because Compton heating timescale is much longer than gas infall timescale. Inside $10^4r_s$, Compton scattering even plays a cooling role, because in this region, gas temperature is much higher than Compton temperature. In model windFB22, outflow is strong in the whole computational domain, the mass inflow rate keeps decreasing from the outer boundary to the inner boundary. Due to the strong outflow in model windFB22, the mass accretion rate at the inner boundary is significantly reduced. Also, gas density in model windFB22 is much smaller than that in model radFB22 (see middle panel of Figure \ref{Fig:density22}). As mentioned above, the wind feedback can also convert some energy of wind to the internal energy of accretion flow. Therefore, gas temperature in model windFB22 is higher than that in model radFB22 (see bottom panel of Figure \ref{Fig:density22}).

We have shown in our previous paper that the outflow in model radFB22 is spherical distributed (see Figure 1 in Bu \& Yang 2018). This is because that in this model, outflow is launched due to Compton heating. Compton heating rate is spherically distributed. Now, we study in the case of wind feedback, what is the geometry of outflow. We plot Figures \ref{Fig:vector22} and \ref{Fig:vectorzoomwindFB22}. In the left panel of Figure \ref{Fig:vector22}, we plot time-averaged (from $t=1.2\times10^5$ to $2.5 \times 10^5$ year) logarithm density (colors); vectors show time-averaged (from $t=1.2\times10^5$ to $2.5 \times 10^5$ year) velocity vector. We see that in the region $30^\circ<\theta<80^\circ$, the gas flows inwards. Outflows are present close to the rotational axis ($\theta < 30^\circ$) and around midplane. The outflow velocity around rotational axis is much higher than that around midplane. The outflow density around midplane is much higher than that in the region close to rotational axis. In the right panel of Figure \ref{Fig:vector22}, colors show time-averaged (from $t=1.2\times10^5$ to $2.5 \times 10^5$ year) logarithm temperature in unit of virial temperature.; vectors show time-averaged (from $t=1.2\times10^5$ to $2.5 \times 10^5$ year) unit velocity vector. It is clear that in the outflow region, gas temperature is higher than virial temperature. Outflows are pushed out by gas pressure gradient force. In the inflow region, gas temperature is much lower than virial temperature. Figure \ref{Fig:vectorzoomwindFB22} is zooming in of right panel of Figure \ref{Fig:vector22}. One portion of injected wind moves outwards in the region $\theta<45^\circ$. The other portion collides with the infall gas. The infall gas becomes outflow after colliding and moves outwards around midplane. The time-averaged flow is very ordered. But, we note that the snapshot of velocity field is very tangled. Figure \ref{Fig:windfluxtheta22} shows the distribution of time-averaged (from $t=1.2\times10^5$ to $2.5 \times 10^5$ year) mass outflow rate with $\theta$ at outer radial boundary. It is clear, the outflow mass flux in the region close to midplane is comparable to that around rotational axis.

The magnitude of fluctuation of accretion rate in model fullFB22 is larger than that in model windFB22 (Figure \ref{Fig:mdot22}). This is because in model fullFB22, both radiation and wind feedbacks exist. In windFB22, only wind feedback is included. The time-averaged mass accretion rate in model fullFB22 is larger than that in model windFB22 by a factor of $\sim 4$. This is a surprising result, because one would expect that with the help of radiation feedback, the mass accretion rate should be much smaller in model fullFB22 than that in model windFB22. The reason for higher mass accretion rate in model fullFB22 is as follows. We find that in model fullFB22, in the region $r < 3 \times 10^4r_s$, gas temperature is higher than $10^8K$. Therefore, in the region $r < 3 \times 10^4r_s$, Compton scattering plays a cooling role. We find that in the region $r < 3 \times 10^4r_s$, gas temperature in model fullFB22 is lower than that of model windFB22. Correspondingly, in this region, the gas density in model fullFB22 is higher than that in model windFB22. The radial infall velocities in these two models are roughly same. Therefore, the black hole mass accretion rate in model fullFB22 is higher than that in model windFB22. The black hole luminosity in model fullFB22 is $\sim 33$ times smaller than that in model noFB22. The time-averaged outflow geometry in model fullFB22 is quite similar as that in model windFB22 shown in Figure \ref{Fig:vector22}. This is because that in model fullFB22, outflow is mainly launched due to wind feedback.

\section{Summary and discussion}
We perform two-dimensional simulations to study slowly rotating low-luminosity hot accretion flow at parsec and sub-parsec scale. The feedback effects of radiation and wind from the central LLAGN are taken into account. We set the black hole mass $M=10^8M_{\odot}$.

We set the gas density at the outer boundary to be $\rho_0=10^{-26}{\rm g/cm^3}$ and$ \rho_0=10^{-22}{\rm g/cm^3}$.
Due to the low gas density, the accretion flow is in hot accretion mode. Due to the low gas density, the length scale of radiation feedback (see Equation (16)) is very larger. Or, in other words, the Thompson scattering optical depth of the accretion flow is very small. Radiation can not effectively deposit its energy to the accretion flow. Radiation feedback plays a very minor role in quenching the black hole accretion.

The typical length scale of wind feedback is much small. Wind can effectively interact with the accretion flow. Due to the wind feedback, the accretion rate strongly oscillates with time. In the accretion flows with $\rho_0=10^{-24} {\rm g/cm^3}$, we find that outflow can be present even without feedback from the central LLAGN. If we consider the wind feedback of the central LLAGN, the mass accretion rate can be decreased by a factor of $\sim 3$ compared to models without wind feedback. The luminosity in the model with both radiation and wind feedback can be smaller than that predicted by the Bondi accretion rate by a factor of $400$.

In the accretion flows with higher density ($\rho_0=10^{-22} {\rm g/cm^3}$), wind power is significantly higher. Therefore, wind feedback is more efficient in suppressing the black hole mass accretion rate. Wind feedback can decrease the black hole mass accretion rate by a factor of $100$. We find that in the model with full feedback (fullFB22), the black hole luminosity is smaller than that predicted by the Bondi formula by a factor of $33$.

Pellegrini (2005) calculates the luminosity for many local universe dim galactic nuclei. In her calculation, the black hole accretion rate is assumed to be equal to the Bondi accretion rate. The radiative efficiency used in that paper is $\eta \propto \dot M$, which is given by the advection-dominated accretion flow model (Narayan \& Yi 1995). In that paper, it is found that the black hole luminosity calculated is significantly higher than observations of many local universe galactic nuclei. In this paper, we find that for the accretion flow self-consistently taking into account wind and radiation feedback, the black hole luminosity can be significantly lower than that predicted by the Bondi formula. The results in this paper may be useful to explain the low-luminosity of galactic nuclei in the local universe.

\section*{Acknowledgments}
This work is supported in part by the National Program on Key Research and Development Project of China (Grant No. 2016YFA0400704),  the Natural Science Foundation of China (grants
11573051, 11633006, 11773053 and 11661161012), the Natural Science
Foundation of Shanghai (grant 16ZR1442200), and the Key
Research Program of Frontier Sciences of CAS (No. QYZDJSSW-
SYS008).  This work made use of the High Performance Computing Resource in the Core
Facility for Advanced Research Computing at Shanghai Astronomical
Observatory.

\end{document}